\documentclass[12pt]{iopart}

\usepackage{iopams}
\usepackage{setstack}
\usepackage{graphicx}
\usepackage{subfig}
\usepackage{booktabs}
\usepackage{float}
\usepackage{cite}
\usepackage{amstext}
\usepackage[labelfont=bf,textfont=it]{caption}
\usepackage{url}
\usepackage{hyperref}

\usepackage{xcolor}
\usepackage{multirow}

\usepackage{setspace}
\begin{document}

\title{Theory-based scaling laws of near and far scrape-off layer widths in single-null L-mode discharges}

\author{M. Giacomin$^1$, A. Stagni$^{1,2}$, P. Ricci$^1$, J. A. Boedo$^3$, J. Horacek$^4$, H. Reimerdes$^1$, and C. K. Tsui$^{1,3}$}
\address{$^1$Ecole Polytechnique F\'{e}d\'{e}rale de Lausanne (EPFL), Swiss Plasma Center (SPC), CH-1015 Lausanne, Switzerland}
\address{$^2$Politecnico di Milano, Via Ponzio 34/3, 20133 Milan, Italy}
\address{$^3$Center for Energy Research (CER), University of California (UCSD), San Diego, La Jolla, California 92093}
\address{$^4$Institute of Plasma Physics of the CAS, Za Slovankou 3, Prague, Czech Republic}
\ead{maurizio.giacomin@epfl.ch}

\begin{abstract}
Theory-based scaling laws of the near and far scrape-off layer (SOL) widths are analytically derived for L-mode diverted tokamak discharges by using a two-fluid model. The near SOL pressure and density decay lengths are obtained by leveraging a balance among the power source, perpendicular turbulent transport across the separatrix, and parallel losses at the vessel wall, while the far SOL pressure and density decay lengths are derived by using a model of intermittent transport mediated by filaments. 
The analytical estimates of the pressure decay length in the near SOL is then compared to the results of three-dimensional, flux-driven, global, two-fluid turbulence simulations of L-mode diverted tokamak plasmas, and validated against experimental measurements taken from an experimental multi-machine database of divertor heat flux profiles, showing in both cases a very good agreement.  
Analogously, the theoretical scaling law for the pressure decay length in the far SOL is compared to simulation results and to experimental measurements in TCV L-mode discharges, pointing out the need of a large multi-machine database for the far SOL decay lengths.  

\end{abstract}
Keywords: plasma turbulence, scrape-off layer width, GBS\\
%\submitto{\NF}

\section{Introduction}

In ITER and future fusion reactors, a significant fraction of the fusion power is expected to cross the separatrix and be transported along the magnetic field lines to the target plates through the narrow region of the scrape-off layer (SOL). Due to technological limits imposed by materials, the peak of the heat flux reaching the tokamak wall must not exceed a value of the order of 10~MW/m$^2$~\cite{loarte2007,pitts2013}. 
Without high volumetric power radiation in the SOL and partial divertor detachment, this limit will be certainly exceeded in ITER~\cite{matthews2007,eich2011,eich2013,chang2017,li2019}. Therefore, understanding the mechanisms that regulate turbulent transport in the SOL and predicting the SOL power decay length is of fundamental importance to determine the operational window for a divertor solution compatible with adeguate core confinement, not only for ITER but also for all future high-performance fusion devices. 
While ITER goal is to operate in H-mode where heat exhaust is most severe, the first campaigns, as well as the start-up and landing phase of the H-mode ITER discharges, will be in L-mode, thus prompting the need to provide theoretical scaling laws of the SOL power decay length in this regime. 
Moreover, understanding the mechanisms that regulate the SOL width in L-mode discharges constitutes the first step towards a theory-based scaling law of the near SOL width in H-mode, which involves a more complex physics. 

As observed experimentally (see e.g. Refs.~\cite{labombard2001,boedo2001,rudakov2005,carralero2017,kuang2019}), the SOL presents two density and power decay lengths; a shorter one in the near SOL and a longer one in the far SOL, which is the result of different turbulence dynamics in these two regions. Indeed, as experimentally shown, e.g., in Refs.~\cite{furno2008,xu2009}, turbulence dynamics is wave-like in the near SOL and intermittent in the far SOL.
As discussed, e.g., in Ref.~\cite{horacek2020}, the power fall-off length at the target plates derived  by fitting the divertor heat flux profiles in the vicinity of the strike point is approximated well by the power fall-off length in the near SOL. On the other hand, the heat and particle flux onto main-chamber first wall depends mainly on the intermittent perpendicular transport occurring in the far SOL~\cite{krasheninnikov2008,carralero2015}.  

Significant experimental efforts were made in the past few years to derive experimental scaling laws of the power fall-off length in the near SOL at the divertor plates in L-mode diverted plasmas (see, e.g., Refs.~\cite{loarte1999,ahn2006,scarabosio2013,sieglin2016,horacek2020}).
% Recently, a nonlinear regression has been carried out on power fall-off length measurements based on L-mode hydrogen and deuterium plasma discharges in AUG~\cite{sieglin2016}, leading to
% \begin{equation}
%     \label{eqn:exp_aug}
%     \lambda_q = (1.45\pm 0.13) B_T^{-0.78}q_{\text{cyl}}^{1.07\pm 0.07} P_\text{SOL}^{-0.14\pm 0.05}\,,
% \end{equation}
% where $\lambda_q$ is the power fall-off length in units of mm, $B_T$ is the toroidal magnetic field in units of T, $P_\text{SOL}$ is the power entering into the SOL in units of MW, and
% \begin{equation}
%     q_\text{cyl} = \frac{\pi a^2 (1+\kappa^2)B_T}{\mu_0 R_0 I_p}\,
% \end{equation}
% is the cylindrical safety factor, with $a$ the tokamak minor radius, $\kappa$ the plasma elongation, and $I_p$ the plasma current. In Eq.~(\ref{eqn:exp_aug}), the exponent of $B_T$ is taken directly from the H-mode scaling of Ref.~\cite{eich2011} since all discharges considered in Ref.~\cite{sieglin2016} have the same toroidal magnetic field. 
% The work in Ref.~\cite{sieglin2016} was later generalized 
Recently, a nonlinear regression has been carried out on a set of power fall-off length measurements from a multi-machine database including the COMPASS, EAST, Alcator C-Mod, MAST, and JET tokamaks~\cite{horacek2020}. By combining five hundred L-mode outer and inner divertor heat flux profiles obtained by Langmuir probes or IR camera, thirteen credible scaling laws were derived. A scaling law in good agreement with the experimental data (describing R$^2$=92\% of data variation) obtained by considering only outer divertor measurements is~\cite{horacek2020}
\begin{equation}
    \label{eqn:horacek}
    \lambda_q = 2800 \Bigl(\frac{a}{R_0}\Bigl)^{1.03}f_{Gw}^{0.48}j_p^{-0.35}\,,
\end{equation}
where $\lambda_q$ is the power fall-off length in units of mm, $f_{Gw}$ is the Greenwald fraction, $a$ is the tokamak minor radius, $R_0$ is the tokamak major radius, and $j_p$ is the plasma current density in units of MA/m$^2$.

In parallel to the experimental effort, recent theoretical and numerical studies based on two-fluid models, justified by the high plasma collisionality of this region, have investigated the mechanisms that regulate the near SOL width in L-mode, leading to analytical and numerical scaling laws of the SOL density and pressure gradient lengths in both limited~\cite{halpern2013,halpern2014,halpern2016,fedorczak2019} and diverted~\cite{militello2013,myra2016,giacomin2020,beadle2020} geometries. 
A direct comparison of theoretical scaling laws to experimental data has been carried out in limited geometry~\cite{halpern2013,horacek2015narrow,halpern2016,nespoli2017comp}, showing a good agreement with experimental measurements. However, no in-depth comparison between first-principles theory-based scaling laws and experimental data taken from a multi-machine database has been carried out in L-mode diverted geometry.

While turbulence in the near SOL is characterized by a wave-like dynamics, turbulent transport in the far SOL is dominated by intermittent events due to coherent plasma filaments, also known as blobs~\cite{ippolito2011}.  
Filaments extend along the parallel direction with a cross section spatially localized on the poloidal plane and their associated density fluctuations have an amplitude even larger than the background density.
Filaments originate near the last-closed flux surface (LCFS) as the results of the nonlinear saturation of interchange-like instabilities, with the density fluctuations sheared apart by the $\mathbf{E}\times\mathbf{B}$ velocity and detached from the main plasma~\cite{furno2008,ippolito2011}.
Then, the vertical charge separation inside the filaments, caused by magnetic gradients and curvature drifts, generates a vertical electric field that, in turns, gives rise to a radial $\textbf{E}\times \textbf{B}$ drift that transports filaments outwards, contributing significantly to the perpendicular transport in the far SOL, flattening the density and pressure profile, and increasing the plasma-wall contact, as experimentally observed in Ref.~\cite{boedo2003}.
Plasma filaments have been experimentally studied in tokamaks~\cite{goncalves2005,tanaka2009,kube2013,walkden2017,tsui2018,bencze2019}, stellarators~\cite{sanchez2003}, reversed field pinch~\cite{spolaore2004}, and basic plasma devices~\cite{antar2001,carter2006,furno2008,katz2008}. 

An analytical theory, referred to as two-region model, based on considering separately the divertor region and the upstream SOL region, that is from the outboard midplane to the divertor entrance where the interchange drive is largest, has been proposed to describe the propagation of filaments in the SOL.
Four regimes of filament motion have been identified, depending on the mechanism responsible for balancing the charge separation driven by the magnetic curvature and gradient drifts~\cite{myra2006}: the sheath connected regime (C$_s$), where the curvature drive is balanced by the current flowing to the sheath; the ideal interchange mode regime (C$_i$), where the ion polarization current due to fanning of the flux surfaces in the divertor region damps the charge separation; the resistive ballooning regime (RB), where the damping of upstream ion polarization current dominates; and the resistive X-point regime (RX), where the parallel current flowing between the upstream and divertor regions is the key damping mechanism. Each damping mechanism results into a different dependence of filament velocity on filament size. The two key parameters that determine the filament regime are the collisionality parameter,
\begin{equation}
\label{eqn:lambda}
    \Lambda = \frac{\nu_{ei} L_{\parallel 1}^2}{\rho_s \Omega_{ce} L_{\parallel 2}}\,,\\
\end{equation}
and the size parameter~\cite{myra2006},
\begin{equation}
\label{eqn:theta}
    \Theta = \biggl(\frac{a_b}{a_*}\biggr)^{5/2}\,,
\end{equation}
where $\nu_{ei}$ is the electron to ion collision frequency, $\rho_s=c_s/\Omega_{ci}$ is the ion sound Larmor radius, with $c_s=\sqrt{T_e/m_i}$ the sound speed and $\Omega_{ci}=eB/m_i$ the ion cyclotron frequency, $L_{\parallel 1}$ is the parallel connection length from upstream to the divertor region entrance, $L_{\parallel 2}$ is the parallel connection length from the divertor region entrance to the target plate, $\Omega_{ce}=eB/m_e$ is the electron cyclotron frequency, $a_b$ is the filament size in the poloidal plane, and $a_*$ is the reference filament size introduced in Ref.~\cite{yu2003} and then redefined in Ref.~\cite{paruta2019} taking into account effects described by the two-region model,
\begin{equation}
    a_*=\rho_s\biggl(\frac{2L_{\parallel 2}^2}{\rho_s R_0}\frac{n_b}{n'}\biggr)^{1/5}\,,   
\end{equation}
with $n_b$ the average of the filament density and $n'$ the density at the near-to-far SOL interface.
The two-region model for filament motion has been extensively validated against experimental results (see, e.g.,~\cite{theiler2009,avino2016,tsui2018}) and verified through numerical simulations. These include recent numerical nonlinear two-dimensional~\cite{bisai2005,militello2012}, three-dimensional single-seed filament~\cite{Dudson2015,shanahan2016}, as well as three-dimensional self-consistently generated filament simulations in realistic geometry~\cite{nespoli2017,paruta2019,beadle2020}, which have shown a good agreement with the two-region model, also in H-mode plasmas~\cite{churchill2017}. Moreover, the work carried out in Ref.~\cite{beadle2020} in double-null geometry has shown that the far SOL density decay length can be described as the result of the transport associated to filaments, whose velocity is described by using the two-region model.  

In the present work, we analytically derive first-principles based scaling laws of the near and far density and pressure decay length for L-mode diverted plasmas. We exploit the results of three-dimensional, flux-driven, global, two-fluid turbulence simulations carried out by using the GBS code~\cite{Ricci2012,Paruta2018} and we consider a balance among sources, perpendicular turbulent transport across the LCFS, and parallel losses at the vessel wall. We focus here on a single-null divertor geometry and we derive scaling laws in terms of engineering parameters, such as tokamak major and minor radii, plasma elongation, toroidal magnetic field at the tokamak axis, edge safety factor, and power crossing the separatrix.
The theoretical scaling laws are then compared to simulation results and against experimental data, showing a very good agreement in the near SOL, while suggesting the need of a large multi-machine database for the far SOL decay lengths.

The paper is organized as follows. The physical model implemented in GBS and used to derive the near and far SOL pressure decay lengths is described in Sec.~\ref{sec:model}. An overview of simulation results is presented in Sec.~\ref{sec:simulations}. In Sec.~\ref{sec:sol_decay}, we derive theoretical scaling laws of the near and far SOL density and pressure decay lengths. A comparison between the analytical scaling laws and experimental results is reported in Sec.~\ref{sec:comparison}. Finally, the conclusions follow in Sec.~\ref{sec:conclusions}.

\section{Simulation model}\label{sec:model}

The model considered in this work is based on the drift-reduced Braginskii two-fluid plasma model implemented in the GBS code \cite{Ricci2012,halpern2016,Paruta2018}. A detailed description of the model can be found in Ref.~\cite{giacomin2020}. 
The use of a drift-reduced fluid model is justified by the high plasma collisionality in the SOL, $\lambda_e \ll L_\parallel \sim 2\pi q R$, with $\lambda_e$ the electron mean-free path, and by the large scale fluctuations, $k_\perp \rho_i \ll 1$, with $\rho_i$ the ion Larmor radius, that dominate transport in this regime.
The model is electrostatic, makes use of the Boussinesq approximation~\cite{Ricci2012,yu2006}, and neglects the interplay between plasma and neutrals, although this is implemented in GBS \cite{wersal2015}. 
The theoretical model proposed in Refs.~\cite{Scott1997,rogers1998} and experimentally studied in Ref.~\cite{labombard2005} suggests an important role of the electromagnetic effects on turbulence regimes in the plasma boundary. On the other hand, the analysis carried out in Ref.~\cite{halpern2013bal} by means of electromagnetic fluid , and in Ref.~\cite{mandell2020}, through electromagnetic gyrokinetic simulations, shows an almost negligible effect of electromagnetic perturbations on SOL turbulence at low and intermediate values of $\beta$, thus justifying the use of an electrostatic model.
The effect of the Boussinesq approximation is discussed in Refs.~\cite{yu2006,bodi2011}, showing that it has a negligible effect on SOL turbulence and equilibrium profiles, although this approximation cannot be taken for granted in general, as shown in Refs.~\cite{ross2019,stegmeir2019}. 
Moreover, as shown in Ref.~\cite{angus2014}, the Boussinesq approximation can affect the dynamics of large filaments, leading, in particular, to an underestimate of the filament velocity.
Since coupling with neutrals dynamics is not considered in the present work, our analysis is restricted to low-density plasma in low-recycling conditions. 
However, the effect of impurity radiation in the divertor region, which is neglected in the present model, can be important, especially in the case of carbon wall. 
Within these approximations, the model equations are
\begin{eqnarray}
\label{eqn:density}
\fl\frac{\partial n}{\partial t} &=& -\frac{\rho_*^{-1}}{B}\bigl[\phi,n\bigr]+\frac{2}{B}\Bigl[C(p_e)-nC(\phi)\Bigr] 
-\nabla_{\parallel}(n v_{\parallel e}) + D_n\nabla_{\perp}^2 n +s_n\, ,\\
\label{eqn:vorticity}
\fl\frac{\partial \omega}{\partial t} &=& -\frac{\rho_*^{-1}}{B}\bigl[\phi,\omega\bigr] - v_{\parallel i}\nabla_\parallel \omega + \frac{B^2}{n}\nabla_{\parallel}j_{\parallel} 
+ \frac{2B}{n}C(p_e + \tau p_i) + D_{\omega}\nabla_\perp^2 \omega\, ,\\
\label{eqn:electron_velocity}
\fl\frac{\partial v_{\parallel e}}{\partial t} &=& -\frac{\rho_*^{-1}}{B}\bigl[\phi,v_{\parallel e}\bigr] - v_{\parallel e}\nabla_\parallel v_{\parallel e} 
+ \frac{m_i}{m_e}\Bigl(\nu j_\parallel+\nabla_\parallel\phi-\frac{1}{n}\nabla_\parallel p_e-0.71\nabla_\parallel T_e\Bigr) \nonumber \\
\fl&+&\frac{4}{3n}\frac{m_i}{m_e}\eta_{0,e}\nabla^2_\parallel v_{\parallel e} + D_{v_{\parallel e}}\nabla_\perp^2 v_{\parallel e}\,, \\
\label{eqn:ion_velocity}
\fl\frac{\partial v_{\parallel i}}{\partial t} &=& -\frac{\rho_*^{-1}}{B}\bigl[\phi,v_{\parallel i}\bigr] - v_{\parallel i}\nabla_\parallel v_{\parallel i} - \frac{1}{n}\nabla_\parallel(p_e+\tau p_i)
+ \frac{4}{3n}\eta_{0,i}\nabla^2_\parallel v_{\parallel i} + D_{v_{\parallel i}}\nabla_\perp^2 v_{\parallel i}\, ,\\
\label{eqn:electron_temperature}
\fl\frac{\partial T_e}{\partial t} &=& -\frac{\rho_*^{-1}}{B}\bigl[\phi,T_e\bigr] - v_{\parallel e}\nabla_\parallel T_e 
+ \frac{2}{3}T_e\Bigl[0.71\nabla_\parallel v_{\parallel i}-1.71\nabla_\parallel v_{\parallel e}
+0.71 (v_{\parallel i}-v_{\parallel e})\frac{\nabla_\parallel n}{n}\Bigr] \nonumber \\
\fl&+& \frac{4}{3}\frac{T_e}{B}\Bigl[\frac{7}{2}C(T_e)+\frac{T_e}{n}C(n)-C(\phi)\Bigr] 
+ \chi_{\parallel e}\nabla_\parallel^2 T_e + D_{T_e}\nabla_\perp^2 T_e + s_{T_e}\,,\\
\label{eqn:ion_temperature}
\fl\frac{\partial T_i}{\partial t} &=& -\frac{\rho_*^{-1}}{B}\bigl[\phi,T_i\bigr] - v_{\parallel i}\nabla_\parallel T_i 
+ \frac{4}{3}\frac{T_i}{B}\Bigl[C(T_e)+\frac{T_e}{n}C(n)-C(\phi)\Bigr] - \frac{10}{3}\tau\frac{T_i}{B}C(T_i) \nonumber \\ 
\fl&+& \frac{2}{3}T_i(v_{\parallel i}-v_{\parallel e})\frac{\nabla_\parallel n}{n} -\frac{2}{3}T_i\nabla_\parallel v_{\parallel e} 
+ \chi_{\parallel i}\nabla_\parallel^2 T_i + D_{T_i}\nabla_\perp^2 T_i + s_{T_i}\,,\\
\label{eqn:poisson}
\fl\nabla_\perp^2\phi &=& \omega-\tau\nabla_\perp^2 T_i\ .
\end{eqnarray}

In Eqs.~(\ref{eqn:density}-\ref{eqn:poisson}) and in the following of the present paper (unless specified otherwise), the density, $n$, the electron temperature, $T_e$, and the ion temperature, $T_i$, are normalized to the reference values $n_0$, $T_{e0}$, and $T_{i0}$. 
The electron and ion parallel velocities, $v_{\parallel e}$ and $v_{\parallel i}$, are normalized to the reference sound speed $c_{s0}=\sqrt{T_{e0}/m_i}$. The norm of the magnetic field, $B$, is normalized to the toroidal magnetic field at the tokamak magnetic axis, $B_T$. Perpendicular lengths are normalized to the ion sound Larmor radius $\rho_{s0}=c_{s0}/\Omega_{ci}$ and parallel lengths are normalized to the tokamak major radius $R_0$. Time is normalized to $R_0/c_{s0}$. The dimensionless parameters appearing in the model equations are the normalized ion sound Larmor radius, $\rho_* = \rho_{s0}/R_0$, the ion to electron reference temperature ratio, $\tau = T_{i0}/T_{e0}$, the normalized electron and ion viscosities, $\eta_{0,e}$ and $\eta_{0,i}$, the normalized electron and ion parallel thermal conductivities, $\chi_{\parallel e}$ and  $\chi_{\parallel i}$, and the normalized Spitzer resistivity, $\nu = e^2n_0R_0/(m_ic_{s0}\sigma_\parallel) = \nu_0 T_e^{-3/2}$, with 
\begin{eqnarray}
\sigma_\parallel &=& \biggl(1.96\frac{n_0 e^2 \tau_e}{m_e}\biggr)n=\biggl(\frac{5.88}{4\sqrt{2\pi}}\frac{(4\pi\epsilon_0)^2}{e^2}\frac{ T_{e0}^{3/2}}{\lambda\sqrt{m_e}}\biggr)T_e^{3/2},\\
\label{eqn:resistivity}
\nu_0&=&\frac{4\sqrt{2\pi}}{5.88}\frac{e^4}{(4\pi\epsilon_0)^2}\frac{\sqrt{m_e}R_0n_0\lambda}{m_i c_{s0} T_{e0}^{3/2}},
\end{eqnarray}
where $\lambda$ is the Coulomb logarithm.

The dimensionless equilibrium magnetic field, $\mathbf{B}=\pm\nabla\varphi+\rho_*\nabla\psi\times\nabla\varphi$, is written in terms of the poloidal flux function $\psi$, which can be an analytical function or can be obtained from an equilibrium reconstruction, with $\varphi$ being the toroidal angle. The plus (minus) sign refers to the direction of the toroidal magnetic field corresponding to the ion-$\nabla B$ drift pointing upwards (downwards).
The spatial operators appearing in Eqs.~(\ref{eqn:density}-\ref{eqn:poisson}) are the $\mathbf{E}\times\mathbf{B}$ convective term $\bigl[g,f\bigr]=\mathbf{b}\ \cdot\ (\nabla g \times \nabla f)$, the curvature operator $C(f)=B\bigl[\nabla \times (\mathbf{b}/B)\bigr]/2\cdot \nabla f$, the parallel gradient operator $\nabla_\parallel f=\mathbf{b}\cdot\nabla f$, and the perpendicular Laplacian operator ${\nabla_\perp^2 f=\nabla\cdot\bigl[(\mathbf{b}\times\nabla f)\times\mathbf{b}\bigr]}$, where $\mathbf{b}=\mathbf{B}/B$ is the unit vector of the magnetic field. 
The numerical diffusion terms, $D_f\nabla_{\perp}^2 f$, are added for numerical stability and they lead to significantly smaller transport than the turbulent processes described by the simulations. 
The differential operators are discretized on a non-field-aligned $(R,\phi,Z)$ cylindrical grid, by means of a fourth-order finite difference scheme~\cite{Paruta2018} ($R$ and $Z$ are the radial, from the tokamak symmetry axis, and vertical directions).
Details on the numerical implementation of the spatial operators are reported in Ref.~\cite{giacomin2020}.

The source terms in the density and temperature equations, $s_n$ and $s_T$, are added to fuel and heat the plasma.
The density and the temperature sources are analytical and toroidally uniform functions of $\psi(R,Z)$, 
\begin{eqnarray}
    \label{eqn:density_source}
    s_n &=& s_{n0} \exp\biggl(-\frac{\bigl(\psi(R,Z)-\psi_{n}\bigr)^2}{\Delta_n^2}\biggr),\\   
    \label{eqn:temperature_source}
    s_T &=& \frac{s_{T0}}{2}\biggl[\tanh\biggl(-\frac{\psi(R,Z)-\psi_{T}}{\Delta_T}\biggr)+1\biggr], 
\end{eqnarray}
where $\psi_n$ and $\psi_T$ are flux surfaces located inside the LCFS. The density source is localized around the flux surface $\psi_n$, close to the separatrix, and mimics the ionization process, while the temperature source extends through the entire core and mimics the ohmic heating. 
Similarly to Ref.~\cite{giacomin2020}, we define $S_n$ and $S_T$ as the total density and temperature source integrated over the area inside the LCFS,
\begin{equation}
    S_n=\int_{A_{\text{LCFS}}} \rho_* s_n(R,Z)\,\mathrm{d}R\mathrm{d}Z
\end{equation}
and
\begin{equation}
    S_T=\int_{A_{\text{LCFS}}} \rho_* s_T(R,Z)\,\mathrm{d}R\mathrm{d}Z\,,
\end{equation}
where the factor $\rho_*$ appears from our normalization choices. 
Analogously, we define the electron pressure source, proportional to the power source, as $S_p=\int_{A_{\text{LCFS}}} \rho_* s_p\,\mathrm{d}R\mathrm{d}Z$, with $s_p=n s_{T_e} + T_e s_n$ and $s_{T_e}$ the electron temperature source.

The simulation domain we consider encompasses the whole tokamak plasma volume to retain the core-edge-SOL interplay~\cite{Fichtmuller1998,Pradalier2017,grenfell2019}, as presented for the first time in Ref.~\cite{giacomin2020turbulence}. The poloidal cross section has a rectangular shape of radial and vertical extension $L_R$ and $L_Z$, respectively. 
For the analysis of the near and far SOL decay lengths, we consider flux-coordinates ($\nabla\psi,\nabla\chi,\nabla\varphi)$, where $\nabla\psi$ denotes the direction orthogonal to the flux surfaces, $\nabla\varphi$ is the toroidal direction, and $\nabla\chi=\nabla\varphi\times\nabla\psi$.

Magnetic pre-sheath boundary conditions, derived in Ref.~\cite{Loizu2012}, are applied at the target plates. Neglecting correction terms linked to radial derivatives of the density and potential at the target plate, these boundary conditions can be expressed as
\begin{eqnarray}
    \label{eqn:boundary_first}
    v_{\parallel i}&=& \pm \sqrt{T_e+\tau T_i},\\
    v_{\parallel e}&=& \pm \sqrt{T_e+\tau T_i}\ \exp\Bigl(\Lambda-\frac{\phi}{T_e}\Bigr),\\
    \partial_Z n &=& \mp \frac{n}{\sqrt{T_e+\tau T_i}}\partial_Z v_{\parallel i},\\
    \partial_Z \phi &=& \mp \frac{T_e}{\sqrt{T_e+\tau T_i}}\partial_Z v_{\parallel i},\\
    \partial_Z T_e &=&\ \partial_Z T_i=\ 0,\\
    \label{eqn:boundary_last}
    \omega &=& -\frac{T_e}{T_e+\tau T_i}\Bigl[\bigl(\partial_Z v_{\parallel i}\bigr)^2\pm\sqrt{T_e+\tau T_i}\,\partial_{ZZ}^2v_{\parallel i}\Bigr], 
\end{eqnarray}
where $\Lambda=3$. The top (bottom) sign refers to the magnetic field pointing towards (away from) the target plate.  
Details on the numerical implementation can be found in Ref.~\cite{Paruta2018}.

\section{Overview of simulation results}\label{sec:simulations}

We focus here on a set of simulations in the L-mode turbulent transport regime. As described in Ref.~\cite{giacomin2020}, where a dedicated analysis has been carried out to identify the turbulent transport regimes in the tokamak edge, a turbulent transport regime appears in two-fluid simulations at low value of heat source and high value of collisionality, where turbulent transport is driven by the resistive ballooning instability. This mode has been associated to the L-mode operational regime of tokamaks.

The simulations are carried out with the following parameters: $\rho_*^{-1}=500$, $a/R_0\simeq 0.3$, $\tau=1$, $\eta_{0,e}=5\times10^{-3}$, $\eta_{0,i}=1$, $\chi_{\parallel e}=\chi_{\parallel i}=1$, $D_f=6$ for all fields, $L_R=600$, $L_Z=800$, $s_{n0}=0.3$, $\Delta_n = 800$, $\Delta_T = 720$,  and different values of $s_{T0}$ and $\nu_0$ (see Eq.~(\ref{eqn:resistivity})), with both favorable and unfavorable ion-$\nabla B$ drift directions being considered (see Tab.~\ref{tab:overview}). 
The magnetic equilibrium, described in Ref.~\cite{giacomin2020}, is analytically obtained by solving the Biot-Savart law in the infinite aspect-ratio limit for a current density with a Gaussian distribution centered at the tokamak magnetic axis, which mimics the plasma current, and an additional current filament outside the simulation domain to produce the X-point. The value of the plasma current and the width of its Gaussian distribution are chosen to have a safety factor $q_0 \simeq 1$ at the tokamak axis and $q_{95} \simeq 4$ at the tokamak edge. 
If we consider as reference density and electron temperature typical values at the separatrix of a TCV L-mode discharge (major radius $R_0 \simeq $ 0.9 m and toroidal magnetic field at the magnetic axis $B_T \simeq$ 1.4 T), i.e. $n_0 \simeq 10^{19}$~m$^{-3}$ and $T_{e0} \simeq 20$~eV, the size of the simulation domain in physical units is $L_R \simeq $ 30 cm, $L_Z \simeq 40$ cm and $R_0 \simeq $ 25 cm, which is approximately a third of the TCV size. 
Regarding the numerical parameters, the number of grid points in the radial, vertical, and toroidal directions are $N_R\times N_Z\times N_\varphi = 240\times 320\times 80$, while the time-step is $2\times 10^{-5}\ R_0/c_{s0}$.\\
We analyse the simulation results after an initial transient, when the simulations reach a global turbulent quasi-steady state resulting from the interplay between the sources in the closed flux surface region, turbulence that transports plasma and heat from the core to the SOL, and the losses at the vessel. The analysis is carried out on a time interval that, expressed in physical units, is of the order of a few milliseconds. 
In the following, we refer to the equilibrium of any quantity $f$ as its time and toroidal average during the quasi-steady state, $\bar{f} = \langle f\rangle_{\varphi,t}$, and to its fluctuating component as $\tilde{f}= f -\bar{f}$.

\begin{table}
    \centering
    \begin{tabular}{ccc}
    \toprule
    $\mathbf{s_{T0}}$ & $\boldsymbol{\nu_0}$ & \textbf{ion-$\mathbf{\nabla B}$ drift}\\
    \midrule
    0.3 & 2.0 & upwards\\
    \hline
    0.3 & 0.9 & upwards\\
    \hline
    0.3 & 0.9 & downwards\\
    \hline
    0.15 & 2.0 & upwards\\
    \hline
    0.15 & 0.9 & upwards\\
    \hline
    0.15 & 0.9 & downwards\\
    \hline
    0.15 & 0.6 & upwards\\
    \hline
    0.075 & 2.0 & upwards\\
    \hline
    0.075 & 0.9 & upwards\\
    \hline
    0.075 & 0.2 & upwards\\
    \hline
    0.075 & 0.2 & downwards\\
    \bottomrule
    \end{tabular}
    \caption{Dimensionless parameters (temperature source strength $s_{T0}$, normalized resistivity $\nu_0$, and ion-$\nabla B$ drift direction) of the set of GBS simulations considered in this work.}
    \label{tab:overview}
\end{table}

\begin{figure}
    \centering
    \subfloat[]{\includegraphics[height=0.2\textheight]{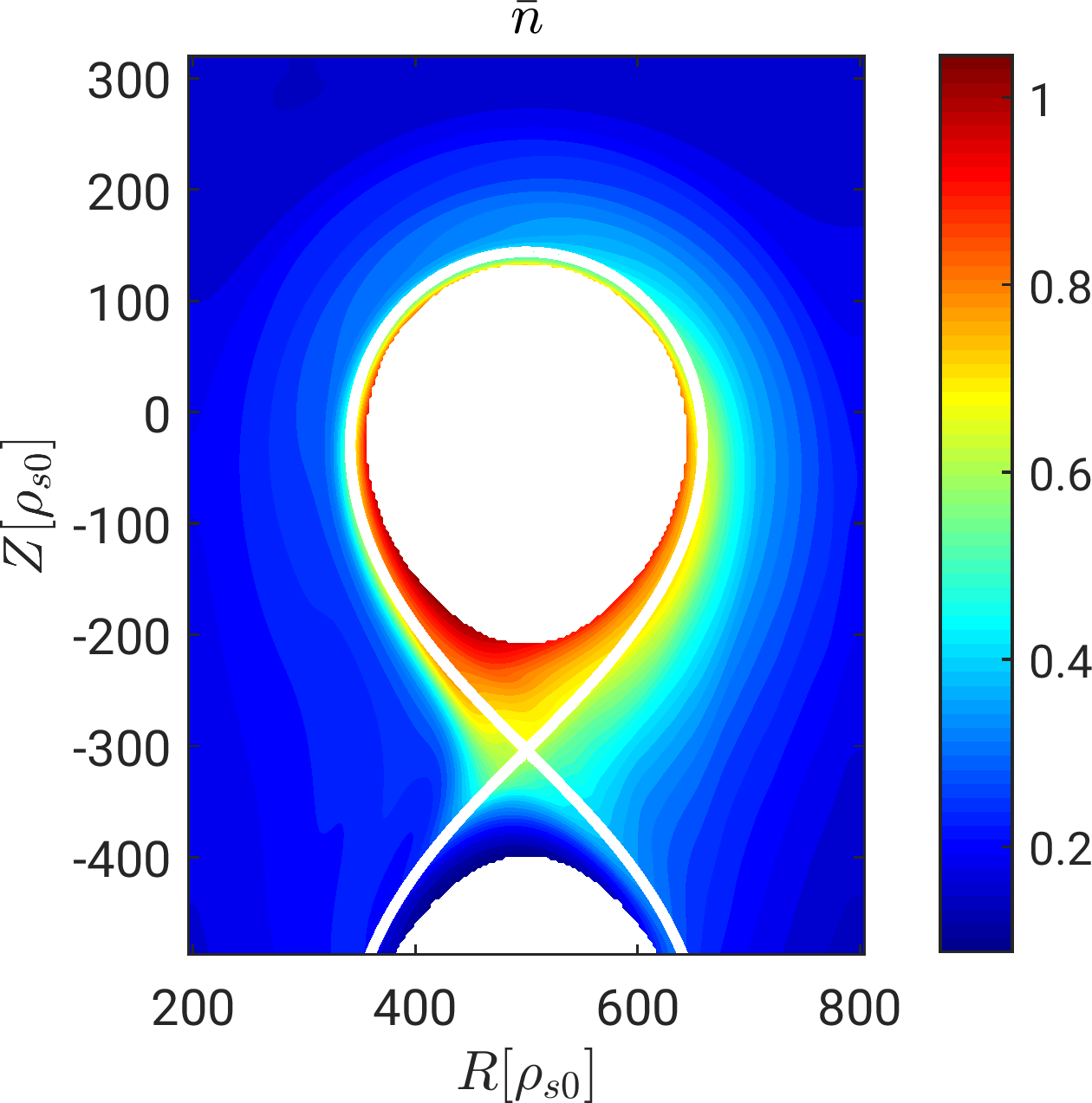}}\quad
    \subfloat[]{\includegraphics[height=0.2\textheight]{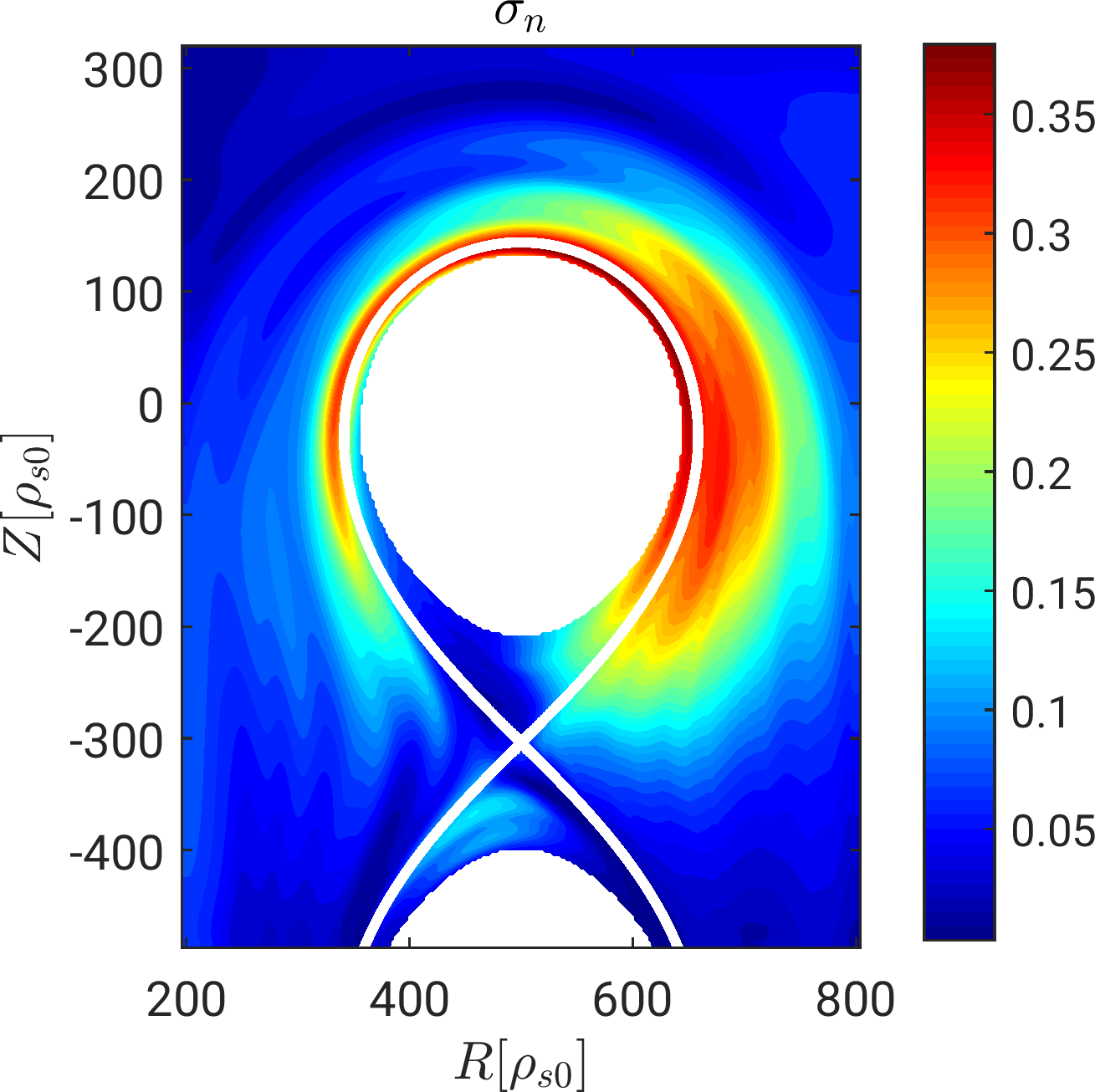}}\quad
    \subfloat[]{\includegraphics[height=0.2\textheight]{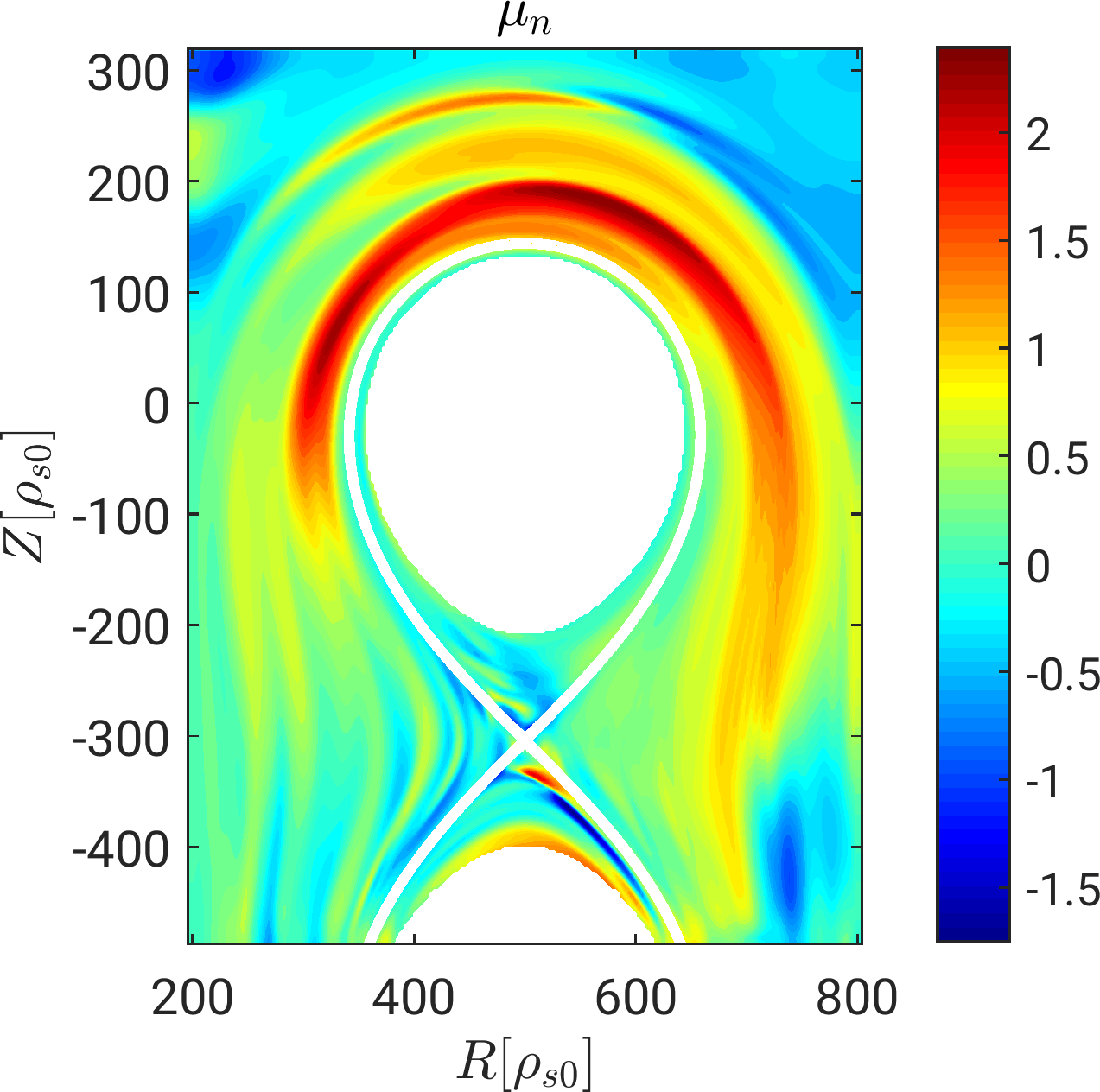}}
    \caption{Equilibrium density, $\bar{n}$, (a), normalized standard deviation, $\sigma_n$, (b) and skewness, $\mu_n$, (c) of density fluctuations at the plasma boundary for the simulation with $s_{T0}$=0.15 and $\nu_0$=0.6. In order to improve the visualisation in the plasma boundary, avoiding the saturation of the colorbar, the core region is not shown. The white line represents the separatrix.}
    \label{fig:overview}
\end{figure}

\begin{figure}
    \centering
    \subfloat[]{\includegraphics[height=0.3\textheight]{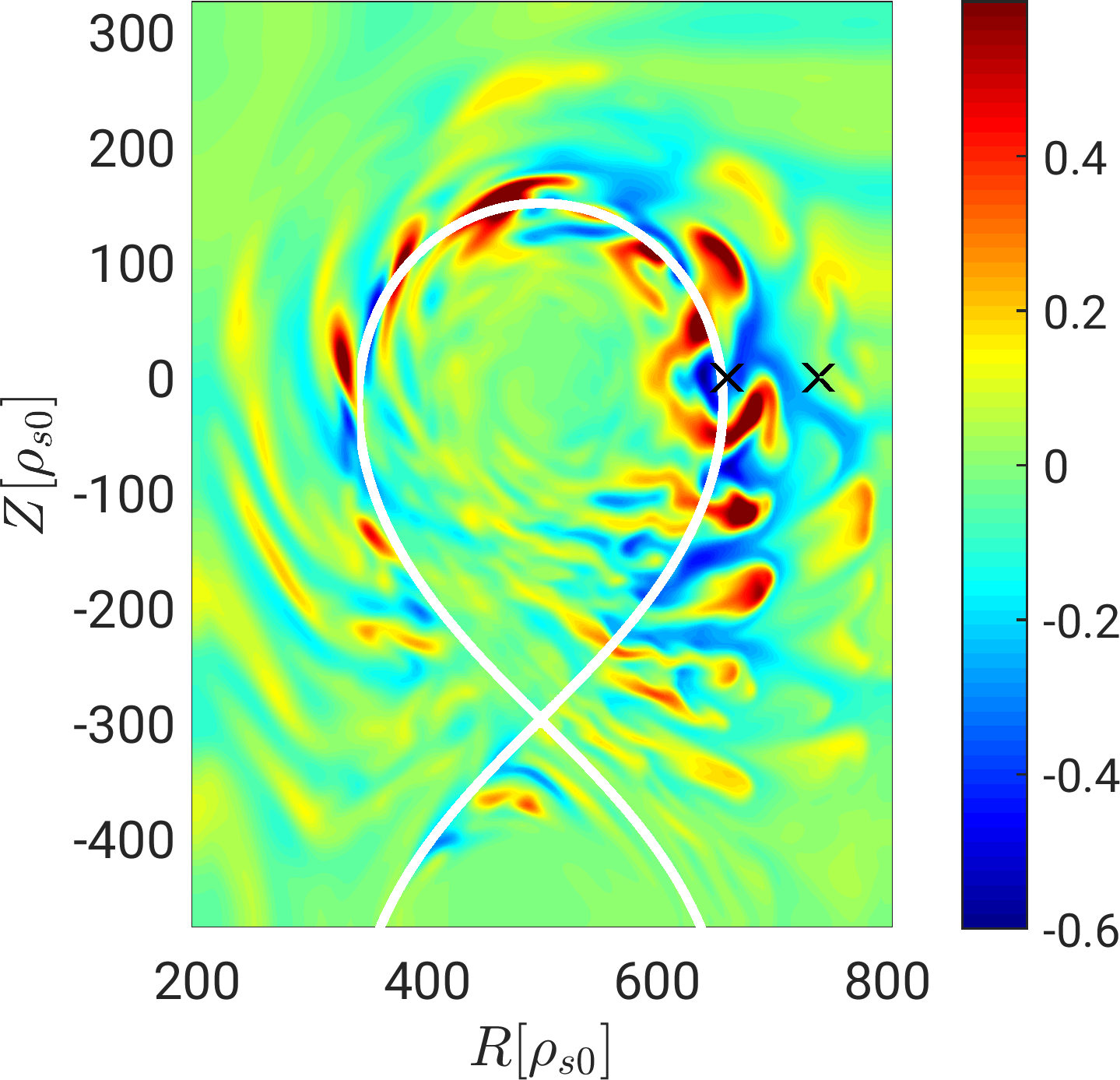}}\qquad
    \subfloat[]{\includegraphics[height=0.3\textheight]{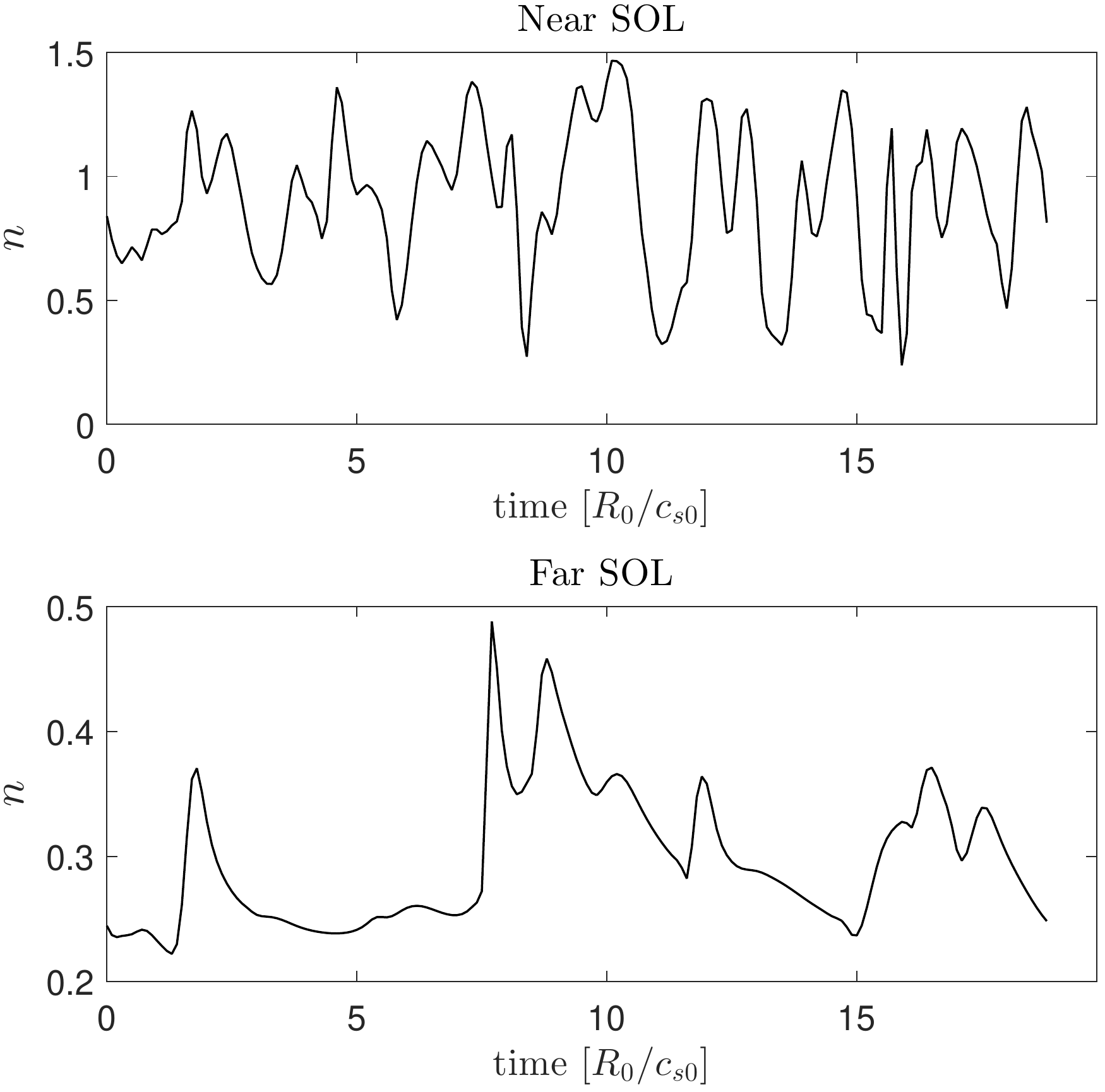}}
    \caption{Typical snapshot of normalized density fluctuations on a poloidal plane (a) and typical time traces of the density in the near and far SOL (b) for the simulation with $s_{T0}$=0.15 and $\nu_0$=0.6. The position where the time traces are extracted is indicated by a black cross.}
    \label{fig:turbulence_near_far}
\end{figure}

As an example of typical simulation results, Fig.~\ref{fig:overview} shows the equilibrium density, the normalized standard deviation and the skewness of density fluctuations in the plasma boundary. The simulation with $s_{T0}$=0.15 and $\nu_0$=0.6 is considered. 
As a consequence of turbulent transport being driven by a resistive ballooning mode~\cite{giacomin2020}, the normalized standard deviation of the density fluctuations peaks on the low-field side (LFS) and remains relatively large throughout the entire LFS SOL, as shown in Fig.~\ref{fig:overview}~(b).
The near and far SOL are characterized by large fluctuations with amplitude comparable to the equilibrium  quantity, as experimentally observed in Refs.~\cite{labombard2001,horacek2005,boedo2009,kube2018}.
The skewness is small in the LFS of the near SOL, suggesting wave-like turbulence, and increases in the far SOL, hinting at the presence of intermittent turbulent events.
A snapshot of the normalized density fluctuations for the same simulation (see Fig.~\ref{fig:turbulence_near_far}~(a)) shows that density fluctuations mainly develop across the separatrix, forming eddies that extend in the radial direction and detach from the main plasma. 
Detached eddies give rise to filaments that radially propagate in the far SOL and are ultimately responsible for its intermittent nature. Therefore, turbulence in the far SOL arises from the steep pressure and density gradients across the separatrix and not from the local equilibrium pressure and density profiles.
The different nature of plasma turbulence in the near and far SOL, namely the highly intermittent and non-local character of turbulence in the far SOL in contrast to wave-like turbulence dynamics in the near SOL, is highlighted in Fig.~\ref{fig:turbulence_near_far}~(b), where two typical time traces of the density in the near and far SOL are shown.

\begin{figure}
    \centering
    \subfloat[]{\includegraphics[width=0.48\textwidth]{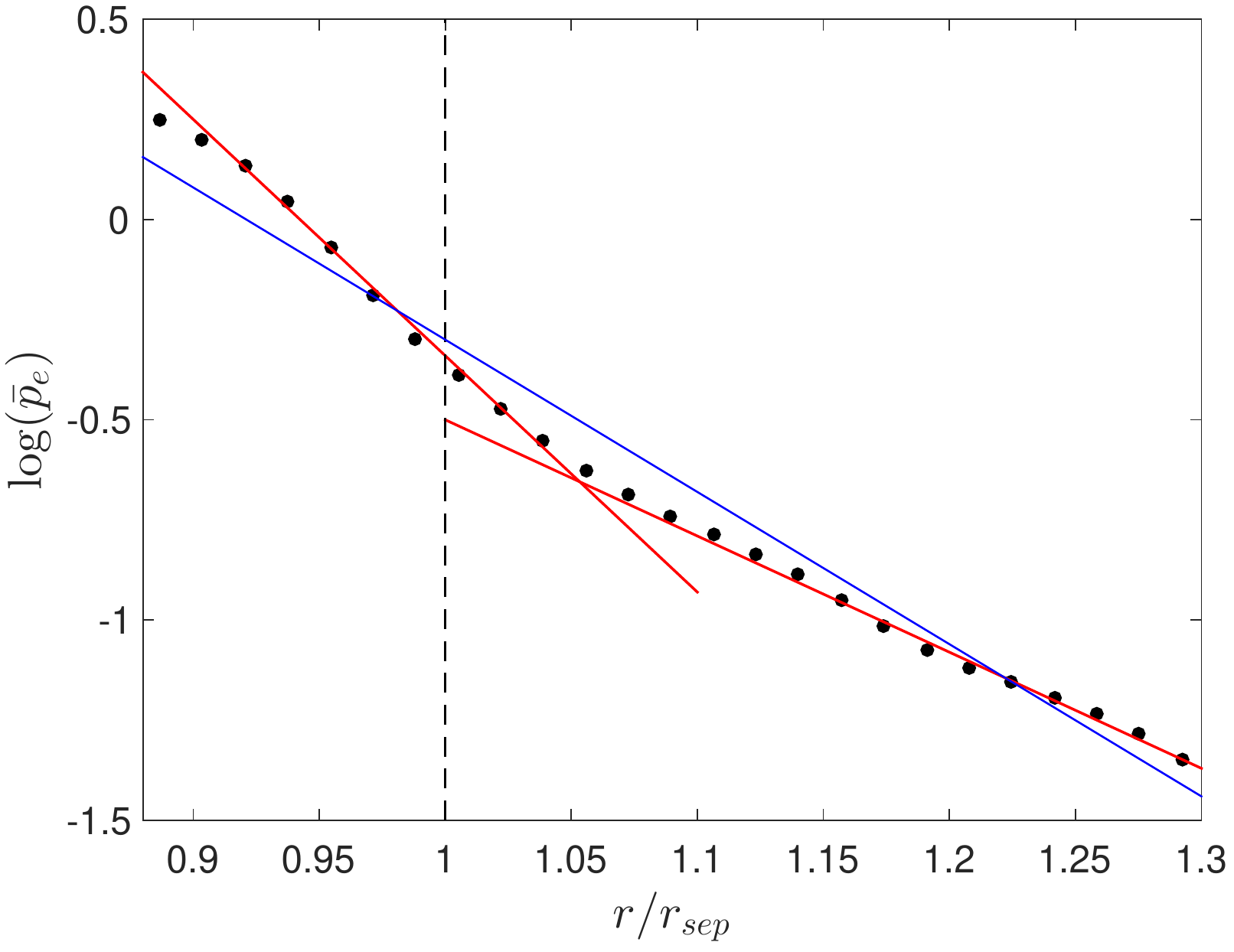}}\quad
    \subfloat[]{\includegraphics[width=0.48\textwidth]{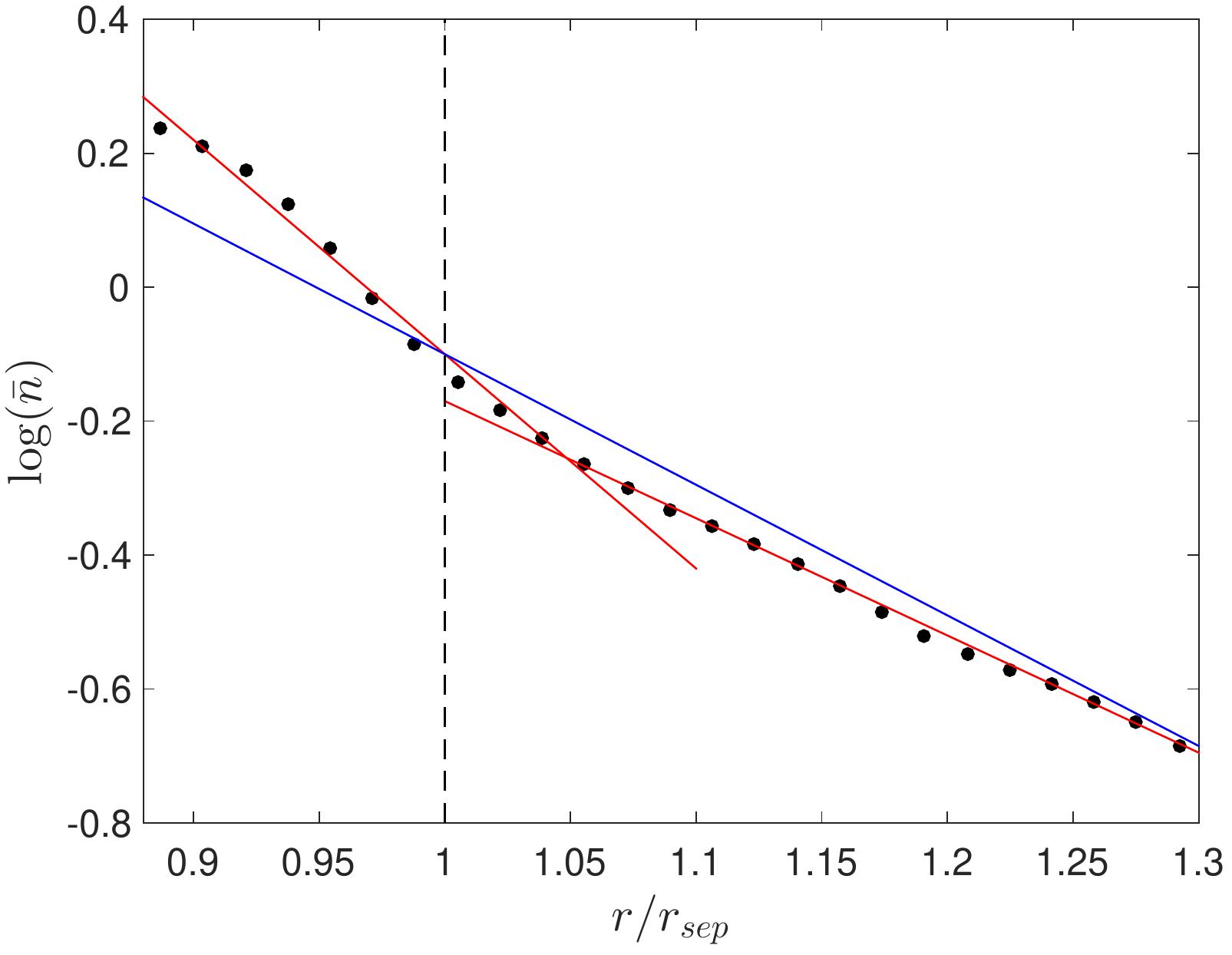}}\quad
    \caption{Radial profile of the equilibrium electron pressure, $\bar{p}_e$, (a) and density, $\bar{n}$, (b) at the outer midplane (black dots) for the simulation with $s_{T0}=0.075$ and $\nu_0=0.2$, and overposed the exponential fits based on one (blue line) or two (red lines) exponential decay lengths. The vertical dashed black line denotes the position of the separatrix.}
    \label{fig:one_two_decay}
\end{figure}

As a consequence of the different transport mechanisms taking place in the near and far SOL, density and pressure show a different decay length in these two regions. 
Two distinct exponential decay lengths have been observed also in experiments (see, e.g., Refs.~\cite{carralero2017,kuang2019}) as well as in other fluid simulations (see, e.g., Refs.~\cite{francisquez2017,beadle2020}). 
An example from one of our simulations is shown in Fig.~\ref{fig:one_two_decay}, where the equilibrium pressure and density radial profiles at the outer midplane are fitted by assuming only one or two distinct exponential decay lengths. 
One exponential overestimates the decay length in the near SOL and underestimates the one in the far SOL. On the other hand, the fit based on two distinct exponential decay lengths agrees well with the equilibrium pressure and density radial profiles on the entire plasma boundary. 
We note that, as revealed by the fit based on two exponential functions, the decay lengths of density and pressure in the near SOL match the ones in the tokamak edge inside the LCFS,  in agreement with experimental observations that show the presence of one characteristic pressure decay length across the separatrix~\cite{sun2015}. 
Since the near-to-far SOL interface radially moves as $\nu_0$ increases, in agreement with experimental observations~\cite{labombard2001,militello2016scrape,kube2018}, the fitting regions in Fig.~\ref{fig:one_two_decay} are properly chosen by identifying the near and far SOL in all the simulations and excluding the transition between these.

\begin{figure}
    \centering
    \subfloat[Pressure decay length in the near SOL.]{\includegraphics[width=0.48\textwidth]{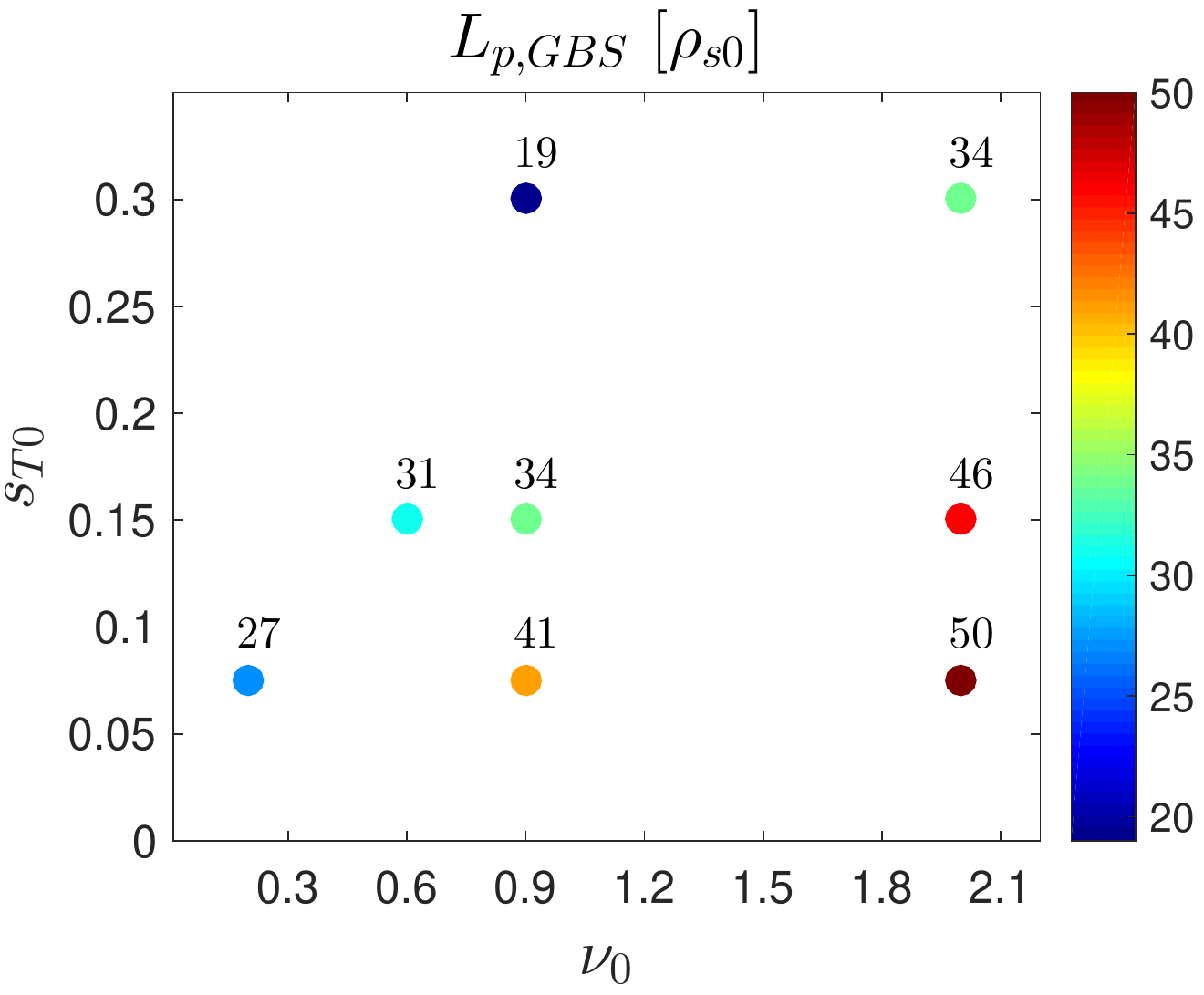}}\quad
    \subfloat[Pressure decay length in the far SOL.]{\includegraphics[width=0.48\textwidth]{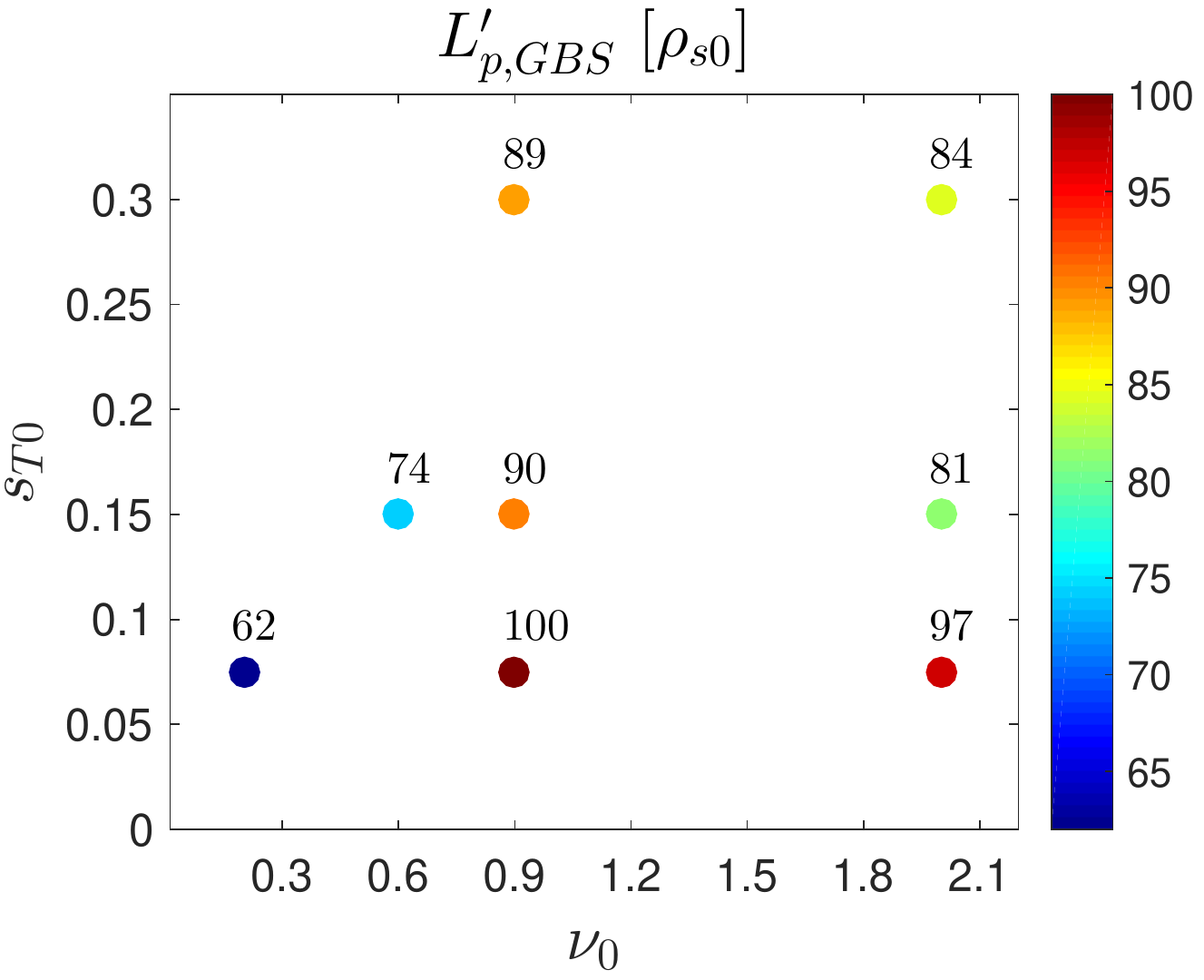}}\\
    \subfloat[Density decay length in the near SOL.]{\includegraphics[width=0.48\textwidth]{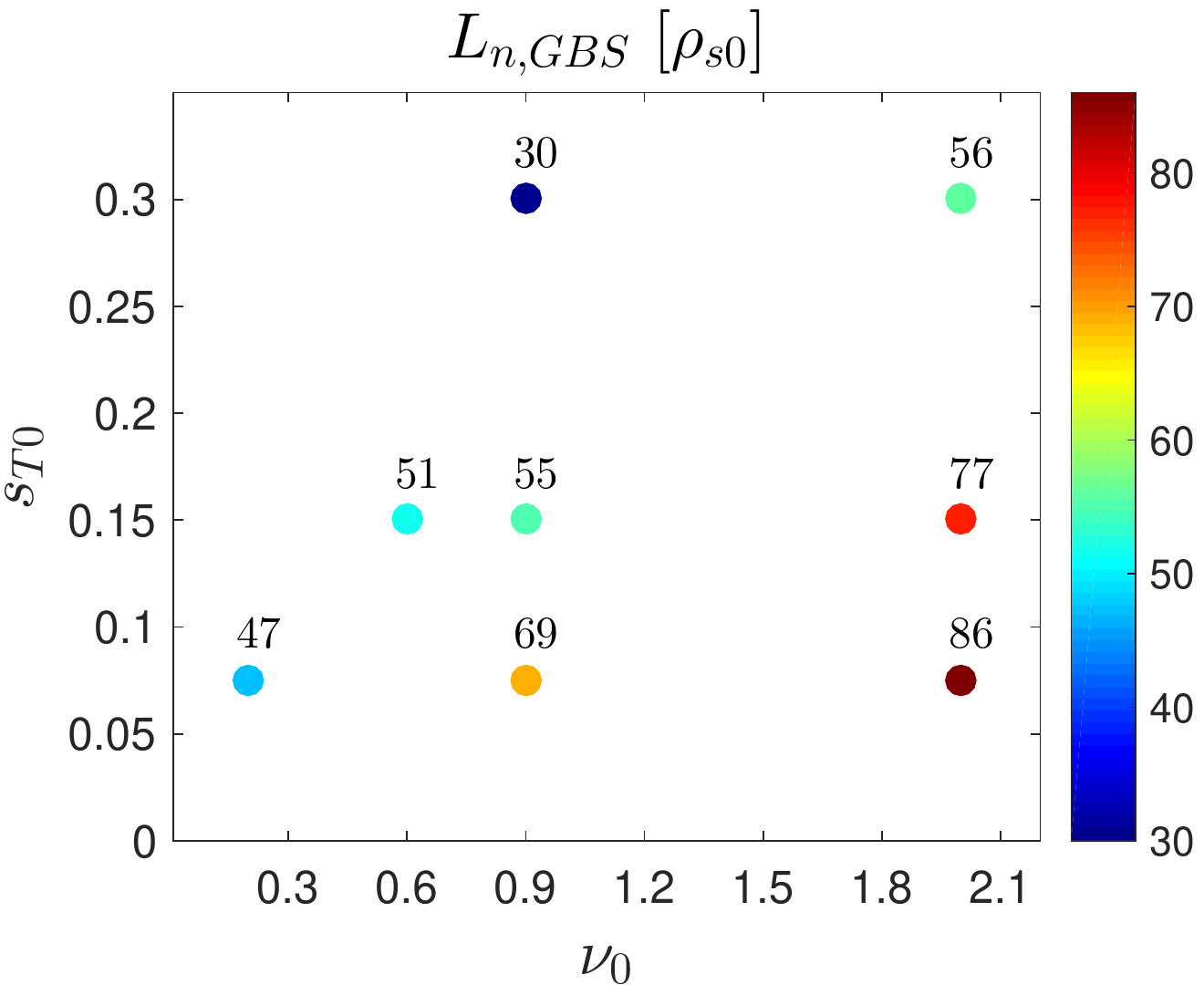}}\quad
    \subfloat[Density decay length in the far SOL.]{\includegraphics[width=0.48\textwidth]{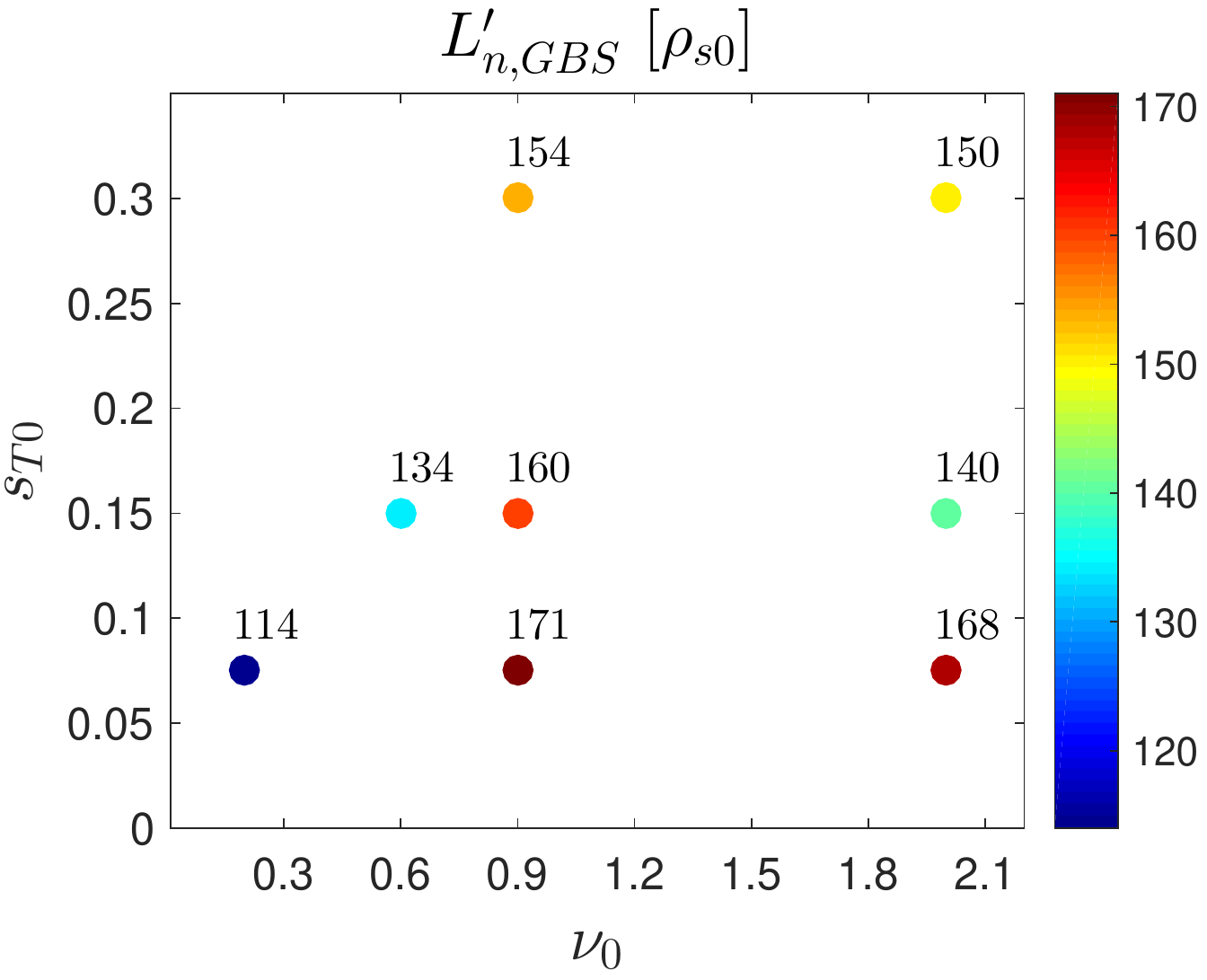}}\\
    \caption{Near and far SOL pressure ((a) and (b)) and density ((c) and (d)) decay lengths, normalised to $\rho_{s0}$, obtained from GBS simulations at the various values of $s_{T0}$ and $\nu_0$ considered in this work with the ion-$\nabla B$ drift pointing upwards. Comparable values are obtained for the simulations with downwards ion-$\nabla B$ drift direction. The numerical value of the equilibrium gradient length is reported next to the data points. }
    \label{fig:near_far_decay}
\end{figure}

Fig.~\ref{fig:near_far_decay} shows the near and far SOL pressure (density) decay lengths, denoted as $L_{p,\text{\tiny{GBS}}}$ ($L_{n,\text{\tiny{GBS}}}$) and $L_{p,\text{\tiny{GBS}}}'$ ($L_{n,\text{\tiny{GBS}}}'$), respectively, for the set of GBS simulations with input parameters listed in table~\ref{tab:overview} (upwards ion-$\nabla B$ drift direction). 
The near SOL pressure gradient length increases as the collisionality increases or the heat source decreases.
In fact, the effective turbulent diffusion coefficient associated to resistive ballooning turbulent transport increases with collisionality and, therefore, for the same value of the input power (i.e. $s_{T0}$), the pressure gradient decreases as $\nu_0$ increases. The detailed description of the $L_p$ dependence on the heat source and plasma collisionality in the resistive ballooning regime is reported in Ref.~\cite{giacomin2020}. 
The far SOL pressure decay length shows a weak dependence on the heat source and collisionality, especially for high values of collisionality and heat source where $L_p'$ seems to saturate.
Similar conclusions can be drawn for the near and far SOL density decay lengths. 

\section{Scaling laws for the near and far SOL decay lengths}\label{sec:sol_decay}

\subsection{Near SOL}

Plasma turbulence in the L-mode transport regime is mainly driven by resistive ballooning modes with the effect of shear flows being negligible, as shown in Ref.~\cite{giacomin2020}. Based on the analysis of the mechanisms that lead to the saturation of the ballooning mode, and a balance between heat sources and turbulent transport, Ref.~\cite{giacomin2020} shows that the equilibrium pressure gradient length in the edge inside the LCFS of the considered L-mode simulations agrees well with the analytical estimate, 
\begin{equation}
\label{eqn:edge_lp}
    L_p\sim \biggl[\frac{\rho_*}{2}(\nu_0 q^2 \bar{n})^2\biggl(\frac{L_\chi}{S_p}\bar{p}_e\biggr)^4\biggr]^{1/3}\,,
\end{equation}
where $L_\chi\simeq 2\pi a\sqrt{(1+\kappa^2)/2}$ is the length of the LCFS poloidal circumference, with $\kappa$ being the plasma elongation, and $\bar{n}$, $\bar{T}_e$, $\bar{p}_e$ are the equilibrium density, electron temperature, electron pressure, which are evaluated at the LCFS~\cite{giacomin2020}. 
Based on the observation (see Fig.~\ref{fig:one_two_decay}) that the exponential decay length of the equilibrium pressure profile in the near SOL and in the edge inside the LCFS correspond, Eq.~(\ref{eqn:edge_lp}) can also be used to estimate the near SOL pressure decay length. 
Indeed, a strong connection between confined edge and near SOL physics has been experimentally observed in Refs.~\cite{brunner2018,faitsch2020,silvagni2020} across various confinement regimes. 

In the present work, we extend the result derived in Ref.~\cite{giacomin2020} by expressing $L_p$ in terms of engineering parameters in order to facilitate both the comparison with experimental results and its applicability to tokamak operation. Therefore, we make explicit in Eq.~(\ref{eqn:edge_lp}) the dependence of $\bar{T}_e$ on $S_p$ and $L_p$, by balancing $S_p$ with the parallel losses at the vessel walls. As an order of magnitude estimate, this balance can be obtained by integrating the heat flux at the vessel wall over the SOL width, 
\begin{equation}
    \label{eqn:sol_balance_first}
    \int_{\text{SOL}} \bar{p}_e \bar{c}_s\, \mathrm{d}l \sim S_p\,.
\end{equation}
In Eq.~(\ref{eqn:sol_balance_first}) we consider the low-recycling regime (i.e. no temperature drop in the divertor region) and we assume that the plasma outflows at the divertor plate with the sound speed.
Moreover, by assuming that the electron pressure and electron temperature decay exponentially in the SOL on the scale $L_p$ and $L_T$, respectively, where $L_T \simeq (1+\eta_e)L_p/\eta_e$, with $\eta_e \simeq 0.77$ derived in Ref.~\cite{Ricci2008} and in agreement with the simulations presented here, $\bar{T}_e$ at the LCFS becomes
\begin{equation}
    \label{eqn:sol_balance}
    \bar{T}_e \sim \biggl[\biggl(1+\frac{\eta_e}{2(1+\eta_e)}\biggr)\frac{S_p}{\bar{n} L_p}\biggr]^{2/3}\,.
\end{equation}

The near SOL estimate of $L_p$ is then derived by replacing $\bar{T}_e$ given by Eq.~(\ref{eqn:sol_balance}) into Eq.~(\ref{eqn:edge_lp}), 
\begin{equation}
    \label{eqn:lp_final}
    L_p \sim \biggl[\frac{1}{8}\biggl(1+\frac{\eta_e}{2(1+\eta_e)}\biggr)^8\rho_*^3\nu_0^6 q^{12} L_\chi^{12}\bar{n}^{10}S_p^{-4}\biggr]^{1/17}\,.
\end{equation}
We note that the equilibrium density gradient length in the near SOL can be directly obtained from Eq.~(\ref{eqn:lp_final}) and the relation $L_n\simeq (1+\eta_e)L_p$ \cite{Ricci2008},
\begin{equation}
    \label{eqn:ln}
    L_n \sim (1+\eta_e)\biggl[\frac{1}{8}\biggl(1+\frac{\eta_e}{2(1+\eta_e)}\biggr)^8\rho_*^3\nu_0^6 q^{12} L_\chi^{12}\bar{n}^{10}S_p^{-4}\biggr]^{1/17}\,.
\end{equation}

\begin{figure}
    \centering
    \subfloat[]{\includegraphics[height=0.25\textheight]{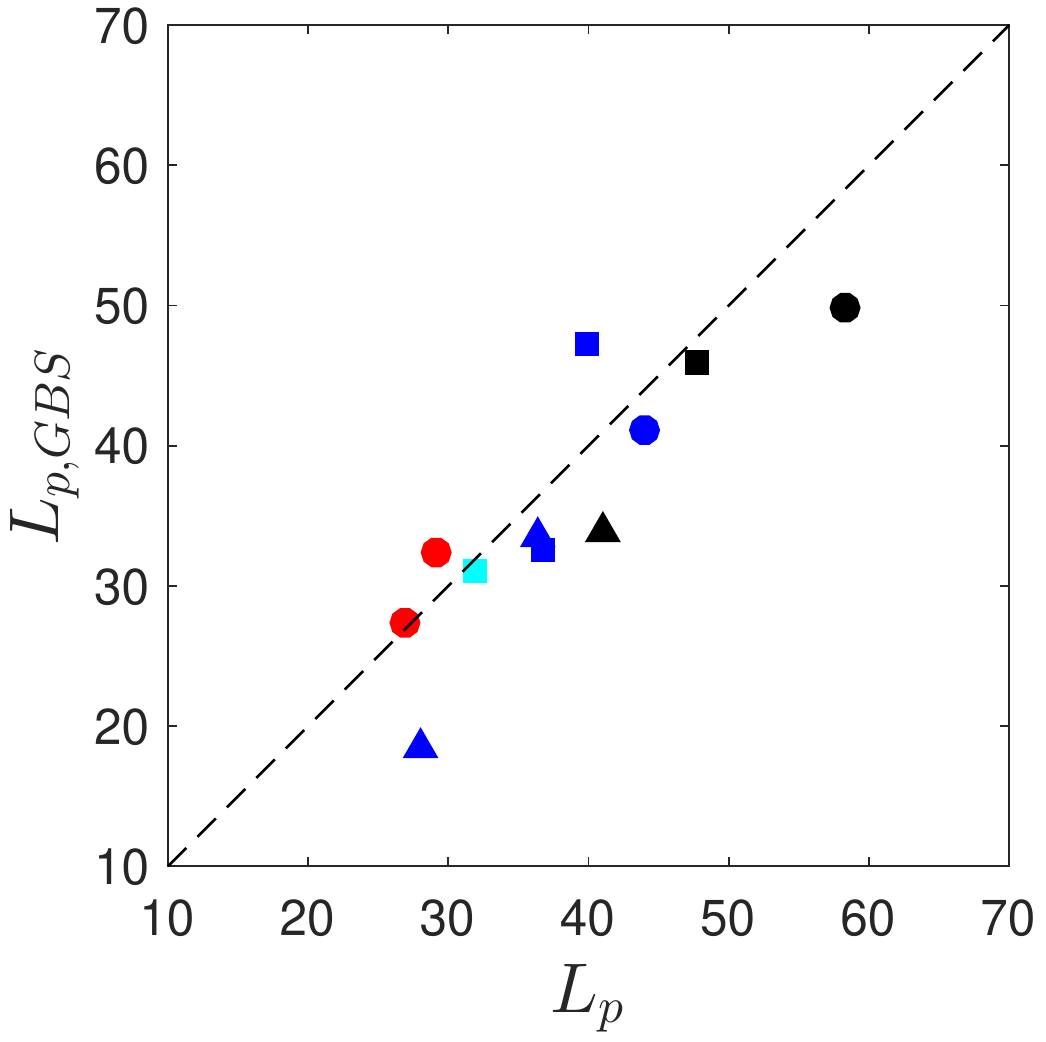}}\quad
    \subfloat[]{\includegraphics[height=0.25\textheight]{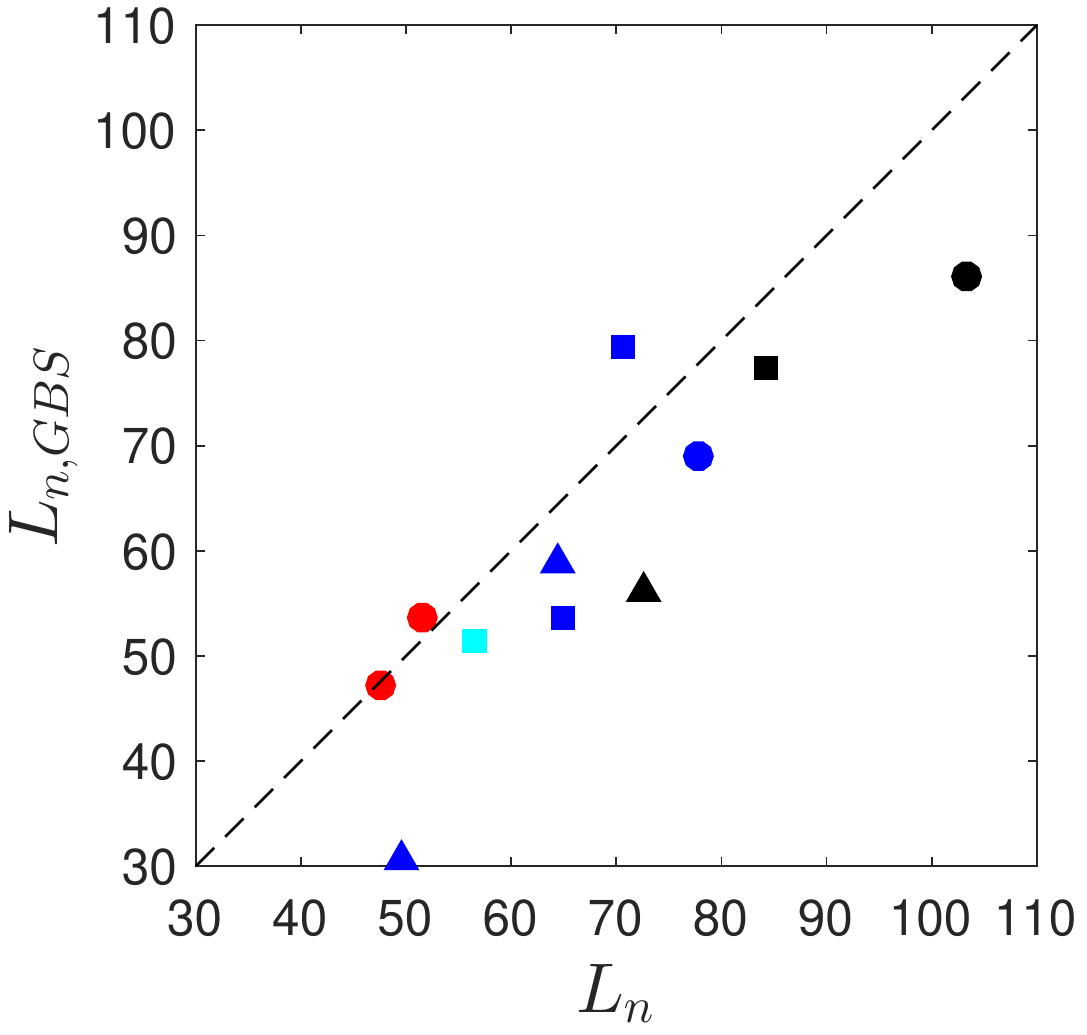}}\quad
    \subfloat{\includegraphics[height=0.25\textheight]{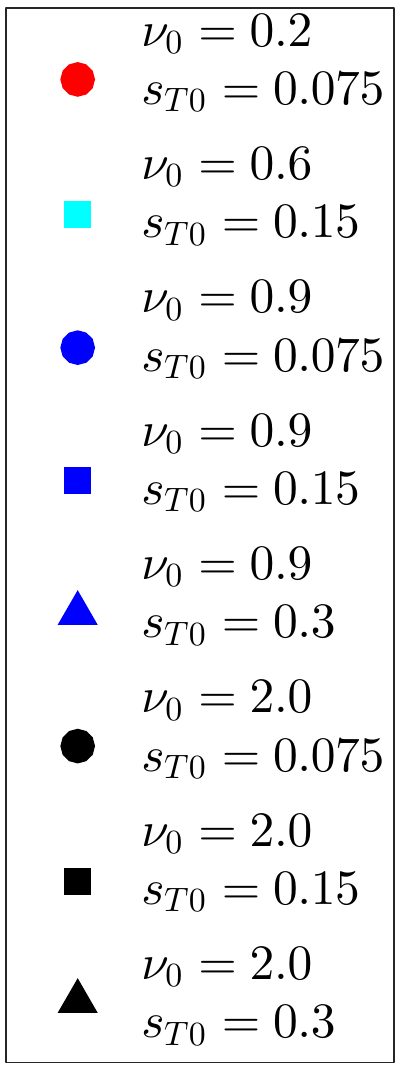}}
    \caption{Comparison between the analytical estimates of the near SOL pressure (a) and density (b) decay lengths and the corresponding ones obtained from GBS simulations.}
    \label{fig:comp_near_sim}
\end{figure}

The theoretical estimates of the near SOL pressure and density decay lengths show a very good agreement with the simulation results for all the values of $s_{T0}$ and $\nu_0$ considered in this work and both directions of the ion-$\nabla B$ drift, as displayed in Fig.~\ref{fig:comp_near_sim}. Indeed, across the set of simulations performed, the difference between simulation results and theoretical predictions is below 20\% for both $L_p$ and $L_n$.

In order to write the analytical scaling law of Eq.~(\ref{eqn:lp_final}) in terms of engineering parameters, such as the power entering into the SOL ($P_\text{SOL}$), the tokamak major and minor radius, and the toroidal magnetic field, we substitute $S_p\sim P_\text{SOL}/(2\pi R_0)$ and $\nu_0$ given by Eq.~(\ref{eqn:resistivity}) into Eq.~(\ref{eqn:lp_final}) and, in physical units, we obtain
\begin{equation}
\label{eqn:lp_phys}
    L_p \simeq 5.6\, A^{1/17} q^{12/17} R_0^{7/17} P_\text{SOL}^{-4/17} a^{12/17}(1+\kappa^2)^{6/17}n_e^{10/17}B_T^{-12/17}\,,
%    5.6\, A^{0.06} q^{0.71} R_0^{0.41} P_\text{SOL}^{-0.24}a^{0.71}(1+\kappa^2)^{0.35}n_e^{0.59}B_T^{-0.71}\,,
\end{equation}
where $L_p$ is in units of mm, $A$ is the mass number of the main plasma ions, $R_0$ and $a$ are in units of m, $P_\text{SOL}$ is in units of MW, $n_e$ is the density at the LCFS in units of $10^{19}$~m$^{-3}$, and $B_T$ is in units of T.

\subsection{Far SOL}

In order to estimate the far SOL density decay length, we consider here a similar approach to the one presented for double-null geometry in Ref.~\cite{beadle2020}. In addition to Ref.~\cite{beadle2020}, we provide also an estimate of the pressure decay length and we write both theoretical scaling laws in terms of engineering parameters.

\begin{figure}
    \centering
    \subfloat[]{\includegraphics[height=0.26\textheight]{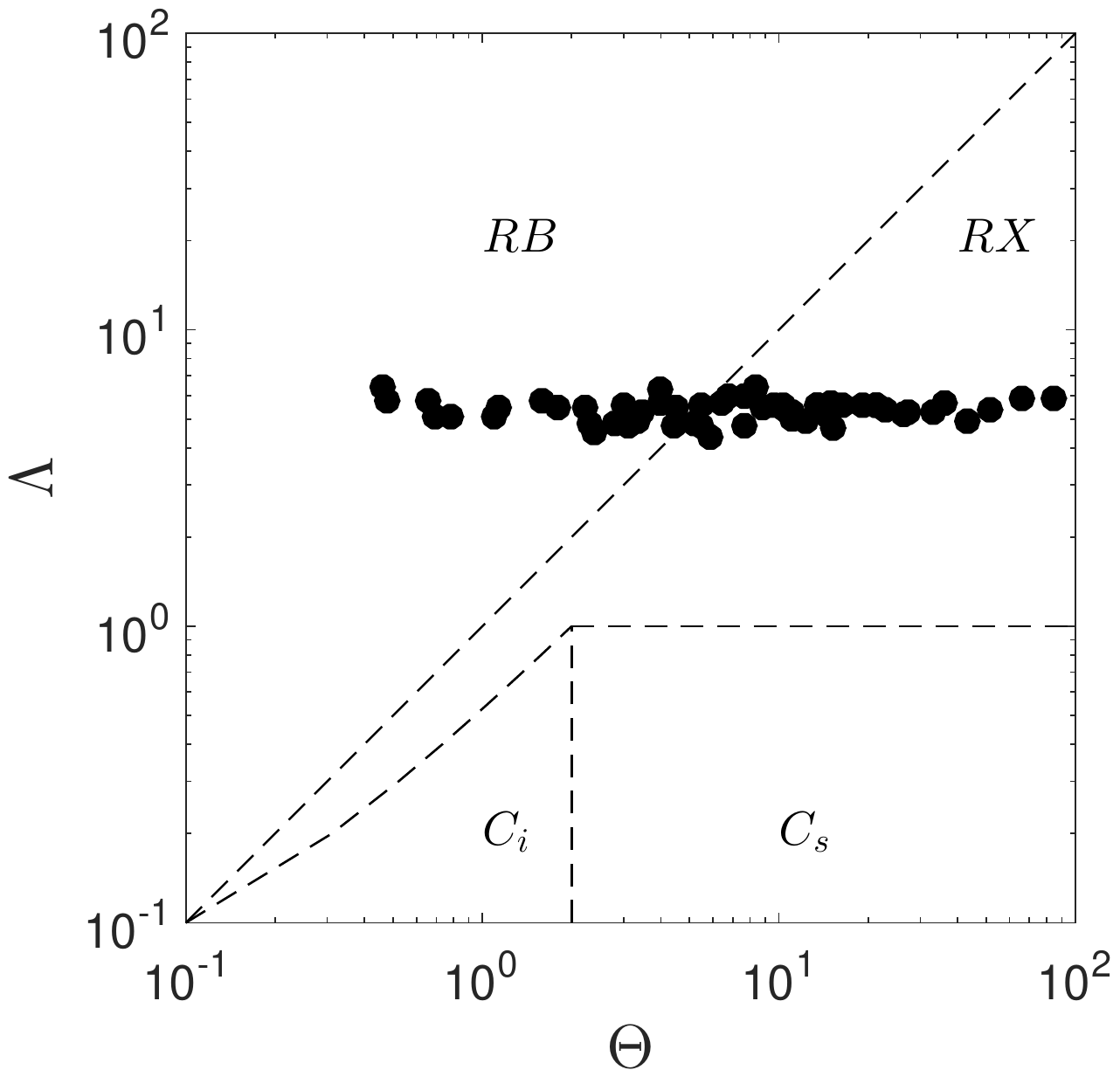}}\,
    \subfloat[]{\includegraphics[height=0.27\textheight]{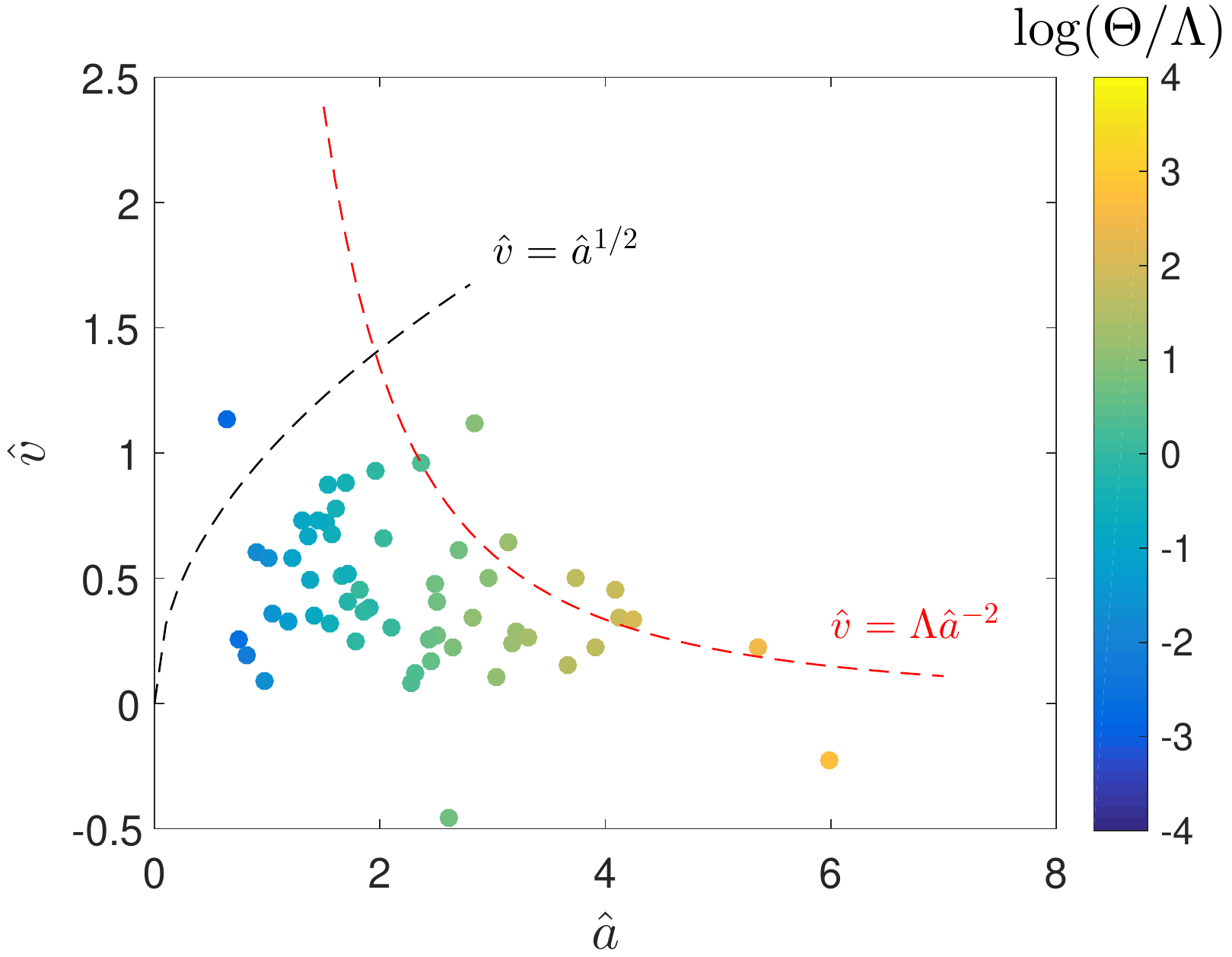}}\\
    \caption{Dispersion plot in the phase space $(\Lambda,\Theta)$ of detected filaments in the simulation with $s_{T0}=0.15$ and $\nu_0 = 0.6$ (a). Black dashed lines are used to delimit the four regimes. Normalized filament velocity as a function of filament size of each detected filament in the same simulation (b). The dashed black line represents the velocity scaling predicted by the two-region model in the RB regime ($\Theta < \Lambda$), while the dashed red line the one in the RX regime ($\Theta > \Lambda$)~\cite{myra2006}. All quantities are obtained by averaging over the filament life.}
    \label{fig:blob}
\end{figure}

As a first step, a pattern-recognition algorithm for filament detection/tracking, described in Refs.~\cite{nespoli2017,paruta2019}, is applied to the GBS simulations considered in this work to determine filament size, velocity, and collisionality parameter, allowing the identification of the filament regime. A typical dispersion plot of the averaged collisionality parameter $\Lambda$, Eq.~(\ref{eqn:lambda}), and size parameter $\Theta$, Eq.~(\ref{eqn:theta}), of each detected filament in the simulation with $s_{T0}=0.15$ and $\nu_0=0.6$ is shown in Fig.~\ref{fig:blob}~(a). The dashed lines delimit the four regimes of filament motion~\cite{myra2006}. 
We note that filaments belong to the RX and RB regimes, a feature in common with all the simulations considered in the present work. We restrict therefore our analysis to the RX and RB regimes.
In Fig.~\ref{fig:blob}~(b), the averaged normalized velocity $\hat{v}=v_b/v_*$ of each filament, with
\begin{equation}
    \label{eqn:blob_ref_vel}
    v_* = c_s\biggl[2\biggl(\frac{\pi a_\psi}{a_\chi}\biggr)^2\frac{n_b}{\bar{n}'}\rho_s^2L_{\parallel 2}\rho_*^2\biggr]^{1/5}
\end{equation}
being the reference filament velocity (see Ref.~\cite{paruta2019}), is displayed for the same simulation as a function of the normalized size, $\hat{a}=a_b/a_*$.
In Eq.~(\ref{eqn:blob_ref_vel}), $a_\psi$ and $a_\chi$ denote the average size of filaments along the $\nabla\psi$ and $\nabla\chi$ direction, respectively.
The normalized filament velocities are mainly scattered between zero and a maximum velocity that varies as a function of size and collisionality in agreement with the analytical normalized velocity predicted by the two-region model~\cite{myra2006}. 
This numerical result agrees with experimental observations that show that the theoretical predictions constitute an upper bound for the filament velocities~\cite{tsui2018}. Indeed, some mechanisms responsible for decreasing the radial filament velocity, such as the filament-filament interaction and the filament rotation, are not included in the two-region model of Ref.~\cite{myra2006}.

Since filament dynamics is responsible of the far SOL pressure and density transport, as shown in Ref.~\cite{beadle2020}, we derive an analytical estimate of equilibrium pressure decay length in the far SOL by balancing the perpendicular transport due to filament motion with parallel heat transport. For this purpose, we take the sum of Eq.~(\ref{eqn:density}), multiplied by $T_e$, and Eq.~(\ref{eqn:electron_temperature}), multiplied by $n$. Then, by time and toroidal averaging the resulting equation, we obtain 
\begin{equation}
\label{eqn:far_start}
    \rho_*^{-1} \partial_\psi \bar{q}_{b,\psi} + \nabla_\parallel (\bar{p}_e\bar{v}_{\parallel e})+\frac{2}{3} 1.71 \bar{p}_e\nabla_\parallel \bar{v}_{\parallel e}+ \frac{2}{3}0.71\bar{T}_e  \bar{v}_{\parallel e} \nabla_\parallel \bar{n} \simeq 0\,.
\end{equation}
In Eq.~(\ref{eqn:far_start}), we identify the perpendicular heat transport with the one mediated by filaments, $\bar{q}_{b,\psi}$.
By assuming that the electron parallel velocity is of the order of $c_s$, and approximating $\partial_\psi  \sim 1/L_p'$ and $\nabla_\parallel \sim 1/L_\parallel$, Eq.~(\ref{eqn:far_start}) yields 
\begin{equation}
    \label{eqn:far_balance}
    \rho_*^{-1}\frac{\bar{q}_{b,\psi}'}{L_p'} \simeq C \frac{\bar{p}_e'\bar{c}_s'}{L_\parallel}\,,
\end{equation}
where $C = 1 + 1.71\,(2/3)+0.71\,(2/3) \simeq 2.6$. In Eq.~(\ref{eqn:far_balance}), the prime symbol appearing in $\bar{p}_e'$, $\bar{c}_s'$, and $\bar{q}_{b,\psi}'$ denotes that these quantities are evaluated at the near-far SOL interface~\cite{paruta2019,beadle2020}. We note that Eq.~(\ref{eqn:far_balance}) only holds in case of negligible variation of electron temperature along the magnetic field lines, which is the case of the low-recycling regime considered here.
The far SOL pressure decay length can then be obtained from Eq.~(\ref{eqn:far_balance}),
\begin{equation}
    \label{eqn:lp_far_first}
    L_p' \sim \frac{\rho_*^{-1}}{C}\frac{\bar{q}_{b,\psi}' L_\parallel}{\bar{p}_e'\bar{c}_s'}\,,
\end{equation}
which relates $L_p'$ to the perpendicular heat flux associated to the filament motion. 

In order to estimate $\bar{q}_{b,\psi}'$, we assume that a filament can be described on the poloidal plane as a coherent structure with Gaussian peak pressure $p_{b,i}$ and half width at half maximum $a_{\psi,i}$, along the $\nabla \psi$ direction, and $a_{\chi,i}$, along the $\nabla \chi$ direction ($i$ is the index identifying the $i$-th filament).
The heat flux associated to the filament motion can be estimated by multiplying the pressure associated with a filament and the filament center of mass radial velocity, $v_{b,i}$, and summing over all the filaments. We obtain
\begin{equation}
    \label{eqn:blob_flux}
    q_{b,\psi}'(\psi,\chi)\sim \sum_i p_{b,i}v_{b,i}\exp\biggl(-\frac{(\psi-\psi_{b,i})^2}{(2a_{\psi,i})^2}-\frac{(\chi-\chi_{b,i})^2}{(2a_{\chi,i})^2}\biggr)\,,
\end{equation}
where $\psi$ and $\chi$ denote coordinate variations along $\nabla\psi$ and $\nabla\chi$, and $(\psi_{b,i},\chi_{b,i})$ are the $i$-th filament center of mass coordinates. 
An estimate of the heat flux due to filament transport is then obtained by averaging $q_{b,i}$ over time and over the LFS SOL area~\cite{russell2007},
\begin{equation}
\label{eqn:total_blob}
\fl\qquad    \bar{q}_{b,\psi}' = \biggl\langle\frac{1}{A_\text{SOL}}\int_{A_\text{SOL}}q_{b}(\psi,\chi)\mathrm{d}\psi\mathrm{d}\chi\biggr\rangle_t = \frac{2\pi}{A_\text{SOL}\log 2}\sum_i\langle a_{\psi,i}a_{\chi,i}p_{b,i}v_{b,i}\rangle_t\,,
\end{equation}
where $A_\text{SOL}$ represents the total far SOL area. 
We note that poloidal variations of the filament size and velocity are present in our simulations, in agreement with experimental observations~\cite{pitts1990,labombard2004}.
By neglecting possible correlation between filaments~\cite{militello2016} and defining $N_b$ as the average number of filaments such that the averaged peak pressure is given by $p_b\sim \sum_i \langle p_{b,i}\rangle_t/N_b$, Eq.~(\ref{eqn:total_blob}) can be approximated as
\begin{equation}
    \label{eqn:total_blob_final}
    \bar{q}_{b,\psi}' \sim \frac{2}{\log 2}f_b p_b v_b\,,
\end{equation}
where $f_b=N_b \pi a_\psi a_\chi /A_\text{SOL}$ is the blob packing fraction. 

In order to make further progress, since filament and background pressure are both progressively drained by the parallel heat flow as moving radially through the far SOL, we assume that the filament peak-to-background pressure ratio remains constant. In Ref.~\cite{beadle2020}, the density fluctuations in the far SOL were assumed to be three times larger than in the near SOL to account for turbulent transport being mainly due to large filaments. Here, as an order of magnitude estimate, we consider that pressure fluctuations in the near and far SOL have similar values, leading to
\begin{equation}
    \frac{p_b}{\bar{p}_e'}\sim \frac{\tilde{p}_e}{\bar{p}_e}\sim \frac{1}{L_p k_\psi}\,.
\end{equation}
The peak filament pressure at the near-far SOL interface can then be obtained,
\begin{equation}
    p_b \sim \frac{\bar{p}_e'}{L_p k_\psi}\,.
\end{equation}
To estimate the blob packing fraction, we evaluate the average filament number by balancing the filament generation and loss rates. As filaments are generated by the nonlinear development of the ballooning instability appearing across the LCFS, the filament generation rate $R_{b,\text{gen}}$ is given by the ballooning mode wavenumber along the LCFS, $L_\chi k_\chi /(2\pi)$, divided by the filament generation time, which can be approximated by the time that a streamer takes to travel its own extension, i.e.  $4 a_\psi/v_b$. We obtain 
\begin{equation}
    \label{eqn:blob_generation}
    R_{b,\text{gen}} \sim \frac{L_\chi k_\chi k_\psi v_b}{4\pi^2}\,.
\end{equation}
The filament loss rate $R_{b,\text{loss}}$ is given by the average filament number on a poloidal plane divided by the time that a filament takes to cross the radial domain,
\begin{equation}
\label{eqn:blob_loss}
    R_{b,\text{loss}} \sim \frac{N_b v_b}{L_\psi}\,.
\end{equation}
The average filament number is then obtained by equating Eqs.~(\ref{eqn:blob_generation})~and~(\ref{eqn:blob_loss}), 
\begin{equation}
\label{eqn:blob_average}
    N_b \sim \frac{A_\text{SOL} k_\chi k_\psi }{4\pi^2}\,,
\end{equation}
where $A_\text{SOL}\simeq L_\chi L_\psi$. By using Eq.~(\ref{eqn:blob_average}), the blob packing fraction becomes
\begin{equation}
    \label{eqn:blob_packing}
    f_b\sim \pi/16\,,
\end{equation}
where we have used $a_\psi\sim \pi/(2k_\psi)$ and $a_\chi\sim \pi/(2k_\chi)$. In all the simulations considered in the present work, the value of $f_b$ is of the order of 0.1 and approximately the same, in agreement with Eq.~(\ref{eqn:blob_packing}) which predicts that $f_b$ is independent of SOL parameters, a feature also observed in experiments~\cite{carralero2018}. We note that the same estimate for $f_b$ is derived in double-null geometry in Ref.~\cite{beadle2020}.

The last quantity to estimate in Eq.~(\ref{eqn:total_blob_final}) is the filament velocity, $v_b=\hat{v}v_*$, where the normalized filament velocity $\hat{v}$ depends on the filament motion regime. From the two-region model (see Refs.~\cite{myra2006,paruta2019} for details), we estimate
\begin{equation}
    \label{eqn:velocity_rb}
    \hat{v}_\text{\tiny{RB}}\sim \hat{a}^{1/2}
\end{equation}
in the RB regime, and
\begin{equation}
    \label{eqn:velocity_rx}
    \hat{v}_\text{\tiny{RX}}\sim \Lambda \hat{a}^{-2}
\end{equation}
in the RX regime, with
\begin{equation}
a_b \simeq \biggl(\frac{2}{\pi}a_\chi\biggr)^{4/5}a_\psi^{1/5}
\end{equation}
and
\begin{equation}
\Lambda = \nu \bar{n}'\frac{L_{\parallel 1}}{c_s L_{\parallel 2}}\,.\\
\end{equation}

By replacing the analytical estimates of $p_b$, $v_b$, and $L_p$ in Eq.~(\ref{eqn:total_blob_final}), the far SOL pressure decay length of Eq.~(\ref{eqn:lp_far_first}) in the RX and RB regimes becomes
\begin{eqnarray}
\label{eqn:far_pres_rx}
\fl\quad L_{p,\text{RX}}'&\sim& \frac{2^{148/85}}{C \log 2}\biggl(1+\frac{\eta_e}{2(1+\eta_e)}\biggr)^{-32/85}(1+\eta_e)^{-3/5}\frac{f_b \bar{n}^{9/17}\nu_0^{27/85}L_\parallel L_{\parallel 1}^2}{q^{116/85}L_\chi^{14/85}S_p^{18/85}\rho_*^{63/85}}\,,\\
\label{eqn:far_pres_rb}
\fl\quad L_{p,\text{RB}}'&\sim& \frac{2^{211/170}\sqrt{\pi}}{C\log{2}}\biggl(1+\frac{\eta_e}{2(1+\eta_e)}\biggr)^{-32/85}(1+\eta_e)^{-1/10}\frac{f_b \bar{n}^{9/17}q^{54/85}\nu_0^{27/85}L_\parallel}{S_p^{18/85}L_\chi^{14/85}\rho_*^{63/85}}\,, 
\end{eqnarray}
where we approximate $\bar{n}'$ and $\bar{T}_e'$ with $\bar{n}$ and $\bar{T}_e$ at the LCFS, with $\bar{T}_e$ given by Eq.~(\ref{eqn:sol_balance}). This approximation is justified by the weak dependence of the ratio $\bar{q}_{b,\psi}'/(\bar{p}_e'\bar{c}_s') \propto v_b/\bar{c}_s'$, appearing in Eq.~(\ref{eqn:lp_far_first}), on the radial position of the near-to-far SOL interface.

The equilibrium density decay length in the far SOL can be obtained by following the same procedure described above for the pressure decay length.  We balance the filament associated perpendicular particle flux, $\Gamma_{b,\psi}$, and the parallel particle transport by considering the leading order terms in Eq.~(\ref{eqn:density}),
\begin{equation}
\label{eqn:den_far_start}
    \rho_*^{-1} \partial_\psi \bar{\Gamma}_{b,\psi} + \nabla_\parallel (\bar{n}\bar{v}_{\parallel e}) \simeq 0\,,
\end{equation}
and we obtain
\begin{equation}
    L_n' \sim \rho_*^{-1} \frac{\bar{\Gamma}_{b,\psi}'L_\parallel}{\bar{n}'\bar{c}_s'}\,,
\end{equation}
where 
\begin{equation}
\label{eqn:far_flux_den}
    \bar{\Gamma}_{b,\psi}' \sim \frac{2}{\log 2}n_b f_b v_b\,,
\end{equation}
with $n_b \sim \bar{n}'/(L_n k_\psi)$. 
By replacing in Eq.~(\ref{eqn:far_flux_den}) the analytical estimates of $n_b$, $v_b$, $L_p$, and $L_n$, the far SOL density decay length in the RX and RB regimes can then be obtained,
\begin{eqnarray}
\label{eqn:far_den_rx}
\fl\quad L_{n,\text{RX}}'&\sim& \frac{2^{148/85}}{\log 2}\biggl(1+\frac{\eta_e}{2(1+\eta_e)}\biggr)^{-32/85}(1+\eta_e)^{-8/5}\frac{f_b \bar{n}^{9/17}\nu_0^{27/85}L_\parallel L_{\parallel 1}^2}{q^{116/85}L_\chi^{14/85}S_p^{18/85}\rho_*^{81/85}}\,,\\
\label{eqn:far_den_rb}
\fl\quad L_{n,\text{RB}}'&\sim& \frac{2^{211/170}\sqrt{\pi}}{\log{2}}\biggl(1+\frac{\eta_e}{2(1+\eta_e)}\biggr)^{-32/85}(1+\eta_e)^{-11/10}\frac{f_b \bar{n}^{9/17}q^{54/85}\nu_0^{27/85}L_\parallel}{S_p^{18/85}L_\chi^{14/85}\rho_*^{81/85}}\,.
\end{eqnarray}

\begin{figure}
    \centering
    \subfloat[]{\includegraphics[height=0.25\textheight]{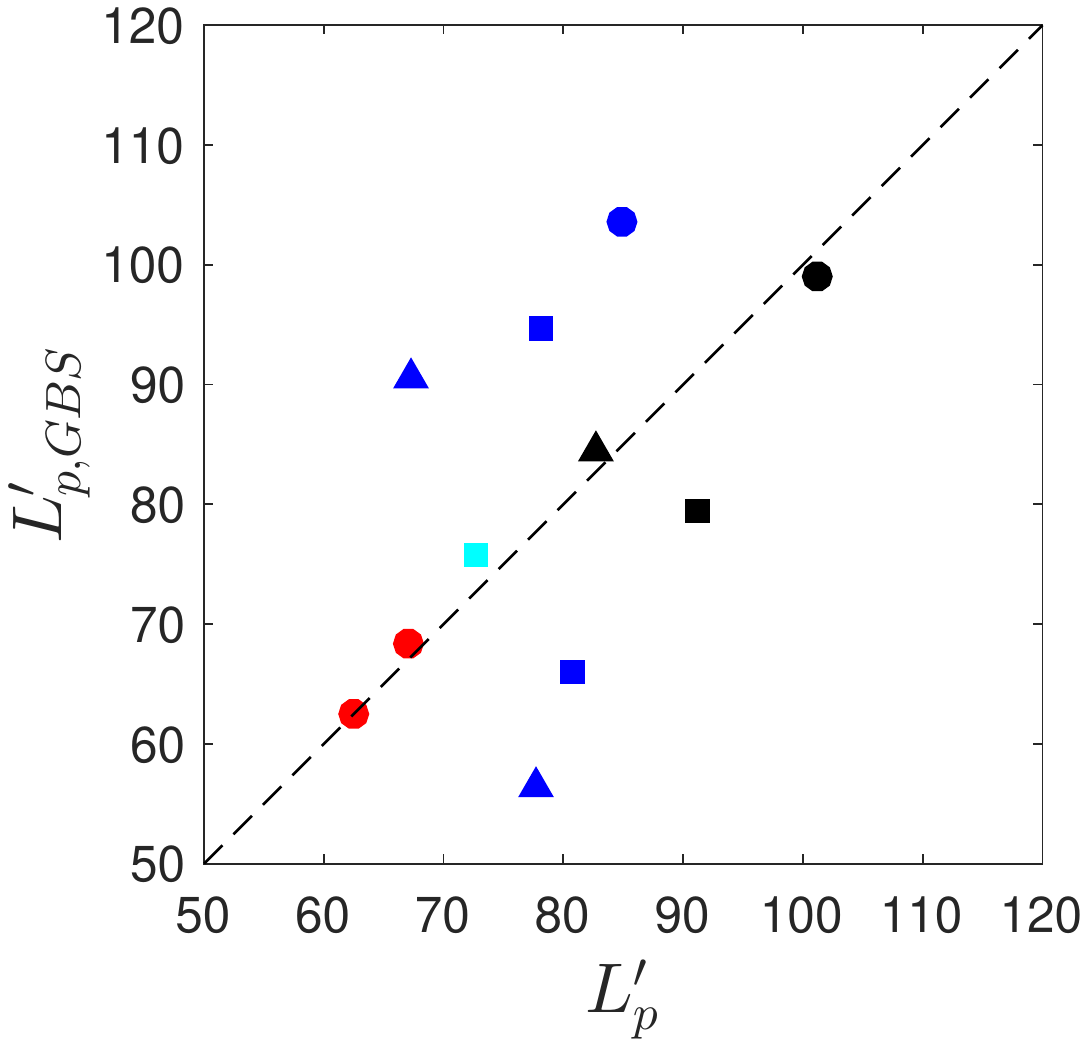}}\quad
    \subfloat[]{\includegraphics[height=0.25\textheight]{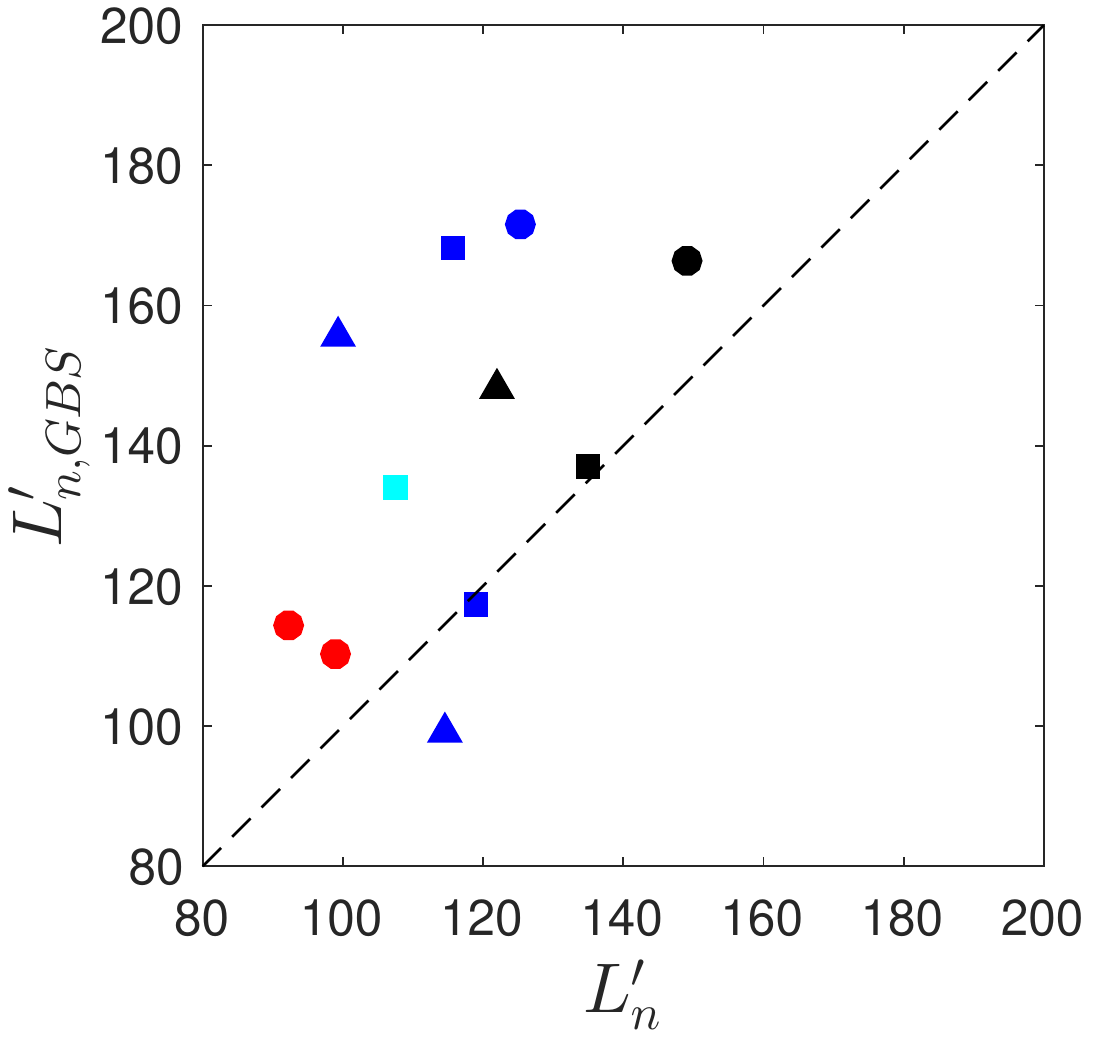}}\quad
    \subfloat{\includegraphics[height=0.25\textheight]{Legend.pdf}}
    \caption{Comparison between the analytical estimates of the far SOL pressure (a) and density (b) decay lengths and the corresponding ones obtained from GBS simulations. 
    The comparison is carried out by numerically solving a filament dispersion relation that links the filament velocity to the filament size (see Ref.~\cite{paruta2019} for details), without considering the limit $\Lambda \gg \Theta$ or $\Lambda \ll \Theta$.}
    \label{fig:comp_far_sim}
\end{figure}

Fig.~\ref{fig:comp_far_sim} shows a comparison between the analytical prediction of the far SOL pressure and density decay lengths and the numerical results obtained from GBS simulations.
This comparison is carried out by numerically solving a filament dispersion relation that links the filament velocity to the filament size (see Ref.~\cite{paruta2019} for details), without considering the limit $\Lambda \gg \Theta$ or $\Lambda \ll \Theta$.
The agreement is good for the pressure and density decay lengths, with differences between theoretical and simulation results of the order of 20\% for the pressure decay length and up to 40\% for the density decay length. 

Similarly to the near SOL decay length, we write the far SOL pressure decay lengths for the RB and RX regimes in terms of engineering parameters. We replace $S_p\sim P_\text{SOL}/(2\pi R_0)$ and $\nu_0$ given by Eq.~(\ref{eqn:resistivity}) into Eqs.~(\ref{eqn:far_pres_rx})~and~(\ref{eqn:far_pres_rb}) and, in physical units, we obtain
\begin{eqnarray}
\label{eqn:far_rx_phys}
\fl\qquad L_{p,\text{RX}}'&\simeq& 3.5 f_b q^{-1.36} R_0^{-1.73} L_\parallel L_{\parallel 1}^2 P_\text{SOL}^{-0.21} a^{-0.16}(1+\kappa^2)^{-0.08}n_e^{0.52}A^{0.05}B_T^{-0.64}\,,\\
\label{eqn:far_rb_phys}
\fl\qquad L_{p,\text{RB}}' &\simeq& 6.1 f_b q^{0.64} R_0^{0.27}L_\parallel P_\text{SOL}^{-0.21} a^{-0.16}(1+\kappa^2)^{-0.08}n_e^{0.52}A^{0.05}B_T^{-0.64}\,,
\end{eqnarray}
where $L_{p,RX}'$, $L_{p,RB}'$ are here in units of mm, $R_0$ and $a$ are the tokamak major and minor radii in units of m, $L_\parallel$ is the parallel connection length from upstream to the outer target plate in units of m, $L_{\parallel 1}$ is the parallel connection length from upstream to the divertor region entrance in units of m, $n_e$ is the density at LCFS in units of 10$^{19}$~m$^{-3}$, $P_\text{SOL}$ is the power entering into the SOL in units of MW, and $B_T$ is the toroidal magnetic field at the magnetic axis in units of T.

\section{Comparison with experimental data}\label{sec:comparison}

\subsection{Near SOL}

We proceed first with the validation of the near SOL pressure decay length derived in Sec.~\ref{sec:sol_decay} against experimental data. For this purpose, we consider the multi-machine database of Ref.~\cite{horacek2020} that contains a set of power fall-off lengths obtained from a nonlinear regression of measurements of divertor heat flux profiles in attached conditions with probes or IR cameras on different tokamaks. Both favorable and unfavorable ion-$\nabla B$ drift directions are considered.  We restrict our comparison to the outer target, considering data from JET, COMPASS, Alcator C-Mod, and MAST tokamaks.  We extend this database by including the TCV $\lambda_q$ measurements in attached conditions presented in Ref.~\cite{maurizio2017}. These values are obtained from heat flux profile measurements at the TCV outer target by using an IR camera.

%The parallel heat flux at the target is related to target quantities via $q_\parallel \sim \gamma_{sh} p_e c_s$, where $\gamma_{sh}$ is the sheath heat transmission coefficient. In order to relate the analytical scaling of $L_p$ at the outboard midplane with $\lambda_q$, we assume that the considered discharges are in the sheath limited regime, where temperature gradients along magnetic field lines can be neglected. Moreover, assuming that $p_e$ and $T_e$ decay exponentially in the SOL on the $L_p$ and $L_T$ scales, respectively, and noting that, in the case considered here, $L_T$ is proportional to $L_p$, the power fall-off length can be written as
%\begin{equation}
%    \label{eqn:power_sol}
%    \lambda_q \sim \biggl(\frac{1}{L_p}+\frac{1}{2L_T}\biggr)^{-1}\propto L_p\,.
%\end{equation}

In order to relate the analytical scaling of $L_p$ at the outboard midplane with $\lambda_q$ experimentally measured at the outer target, we first report $\lambda_q$ upstream accounting for the flux expansion. We also assume that, being the considered discharges in attached conditions, the pressure gradients along the magnetic field lines can be neglected. 
This allows for a direct comparison between $L_p$ in Eq.~(\ref{eqn:lp_phys}) and the experimental $\lambda_q$, i.e. $\lambda_q \propto L_p$, where the proportionality factor is determined from the best fit of experimental and theoretical results, similarly to the procedure outlined in Ref.~\cite{halpern2016}. 
Since only the line-averaged density $\bar{n}_e$ is available in the considered database, we assume the edge density contained in the analytical scaling to be proportional to the line-averaged density, $n_e \propto \bar{n}_e$, where the proportionality factor is included in the unique fitting parameter. 
This assumption is supported by experimental observations that show the presence of an almost linear proportionality between $n_e$ and $\bar{n}_e$ in low-density discharges (see, e.g., Ref.~\cite{ahn2006}). 
The quality of the fit is then expressed through the $R^2$ parameter.

\begin{figure}
    \centering
    \includegraphics[scale=0.55]{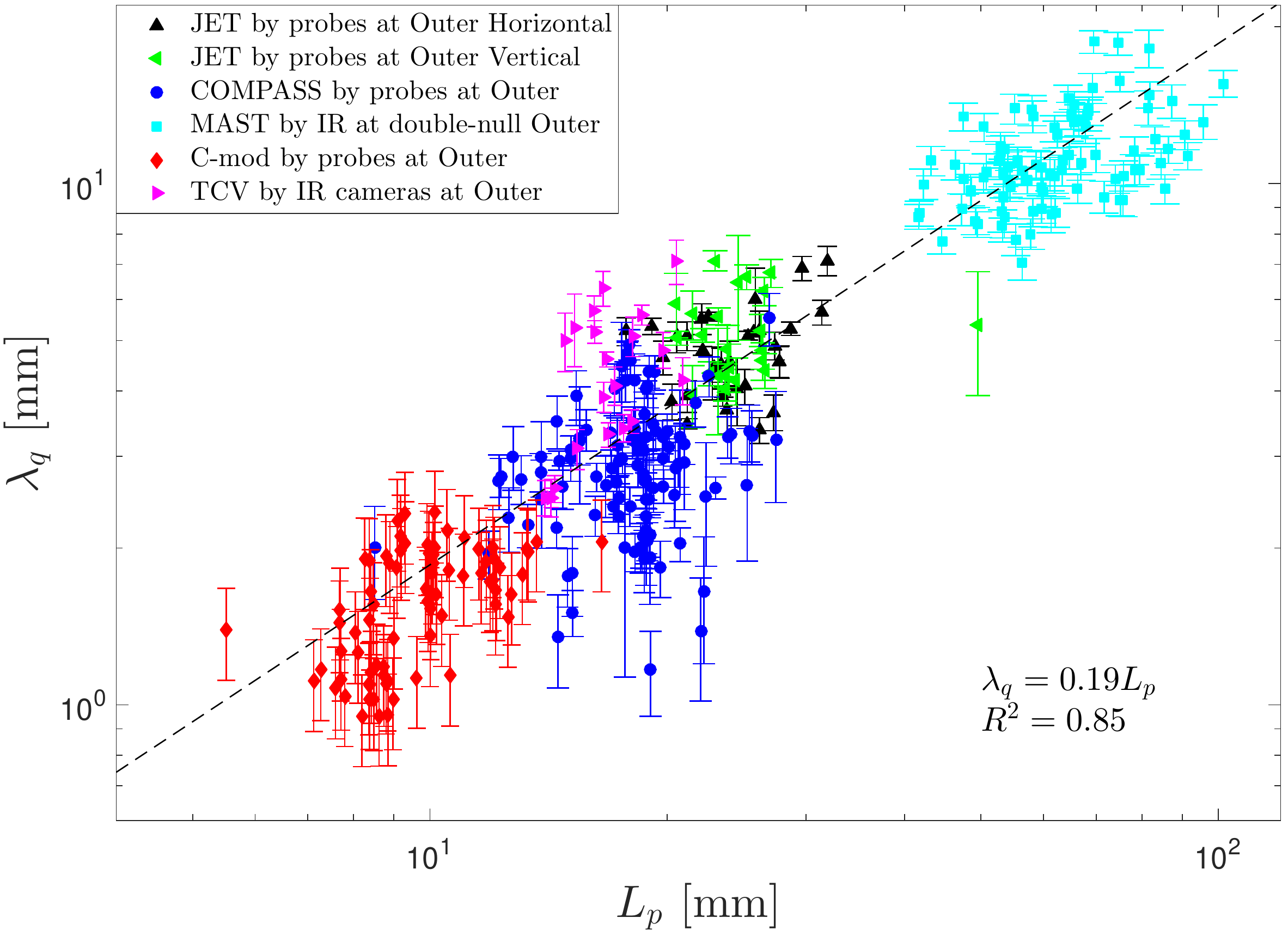}
    \caption{Comparison of the theoretical scaling law of Eq.~(\ref{eqn:lp_phys}) to experimental values of $\lambda_q$ taken from the multi-machine database of Ref.~\cite{horacek2020} extended including TCV data from Ref.~\cite{maurizio2017}. The dashed black line represents the best fit $\lambda_q = \alpha L_p$ with $\alpha$ the unique fitting parameter.}
    \label{fig:comp_exp_near}
\end{figure}

The result of the fitting procedure is shown in Fig.~\ref{fig:comp_exp_near}. The theoretical scaling reproduces experimental data with a very high goodness parameter, $R^2 \simeq 0.85$. 
We highlight that the value of $R^2$ obtained from the comparison between the theoretical scaling and experimental data is even higher than some of the most credible scaling laws derived in Ref.~\cite{horacek2020} from a direct nonlinear regression of experimental results. Indeed, as extensively discussed in Refs.~\cite{horacek2016,horacek2020}, the number of parameters that can be included in a scaling based on the direct nonlinear regression of experimental measurements is limited by their mutual correlation.
For instance, a very strong correlation is found between $R_0$ and $P_\text{SOL}/V$~\cite{horacek2016}, with the consequence that including both of them in the nonlinear regression leads to an ambiguity on their exponent. The mutual correlation between experimental input parameters limits the use of nonlinear regressions to find scaling laws directly from experimental databases, a limitation that is overcome by theory-based first-principles scaling laws, such as the one derived in the present work.

The proportionality constant returned by the fit is approximately 0.2. We note that this constant includes both the proportionality factor between $\lambda_q$ and $\lambda_p$ (we use here $\lambda_p$ to refer to the experimental value of the pressure decay length, while $L_p$ is used for the theoretical prediction of Eq.~(\ref{eqn:lp_phys}))  and the one between $n_e$ and $\bar{n}_e$. By assuming that $p_e$ and $T_e$ decay exponentially in the SOL on the $\lambda_p$ and $\lambda_T \sim 2 \lambda_p$ scales, respectively, the power fall-off length can be written as
\begin{equation}
    \label{eqn:power_sol}
    \lambda_q \sim \biggl(\frac{1}{\lambda_p}+\frac{1}{2\lambda_T}\biggr)^{-1}\sim \frac{4}{5} \lambda_p\,.
\end{equation}
Moreover, from the experimental results shown in Ref.~\cite{ahn2006}, we assume $\bar{n}_e\sim 4 n_e$, which leads to 
\begin{equation}
    \frac{\lambda_q}{L_p}\sim \frac{\lambda_q}{\lambda_p}\Bigl(\frac{n_e}{\bar{n}_e}\Bigr)^{10/17}\sim \frac{4}{5}\Bigl(\frac{1}{4}\Bigr)^{10/17}\sim 0.3\,,
\end{equation}
which is close to the proportionality factor returned by the best fit.

Despite the very high value of $R^2$, we note a dispersion of the experimental measurements around the best fit in Fig.~\ref{fig:comp_near_sim}. This may suggest incomplete or missing dependencies in the theoretical scaling law of Eq.~(\ref{eqn:lp_phys}). In particular, our theoretical scaling law does not include the effect of plasma triangularity, which has been studied with GBS in Refs.~\cite{riva2017,riva2020} for a limited configuration, showing that the near SOL width is enhanced (reduced) by positive (negative) values of triangularity, in agreement with experimental observations~\cite{faitsch2018}. In addition, interchange-like turbulence, which can develop along the divertor leg, can increase the power fall-off length at the target plate~\cite{gallo2017}. This may be especially the case in TCV, where magnetic configurations with a long outer divertor leg are considered. This effect is not included in the present model.

As a further comparison between theoretical and experimental results, we analyse the analogies and differences between the theoretical scaling law of Eq.~(\ref{eqn:lp_phys}) to the one based on the fit of experimental results, reported in Eq.~(\ref{eqn:horacek}).
For this purpose, we rewrite Eq.~(\ref{eqn:lp_phys}) to make explicit the dependence on $f_{Gw}$ and $j_p$,
\begin{equation}
\fl\quad L_p \simeq 8.2 A^{0.06}\Bigl(\frac{n_e}{\bar{n}_e}\Bigr)^{0.59}R_0^{-0.06}\Bigl(\frac{a}{R_0}\Bigr)^{0.47}(1+\kappa^2)^{0.98}\kappa^{-0.12}j_p^{-0.12}\Bigl(\frac{P_\text{SOL}}{S_\text{LCFS}}\Bigr)^{-0.24}f_{Gw}^{0.59}\,,
\end{equation}
where $n_e/\bar{n}_e$ is the ratio of the edge density to the line-averaged density that appears from the definition of $f_{Gw}$ and $S_\text{LCFS}\simeq 4\pi^2 a R_0\sqrt{(1+\kappa^2)/2}$ is the area of the LCFS. We note that the theoretical $L_p$ increases with the aspect ratio and the Greenwald fraction and decreases with the plasma current density, with exponents that are comparable to the experimental ones (see Eq.~(\ref{eqn:horacek})). According to the theoretical scaling, $L_p$ decreases with $P_\text{SOL}/S_\text{LCFS}$, a dependence that is not present in the experimental scaling of Eq.~(\ref{eqn:horacek}), although a similar dependence on $P_\text{SOL}/S_\text{LCFS}$ has been retrieved in other credible experimental scaling laws derived from the same database in Ref.~\cite{horacek2020}. No dependence on $A$ is found in the experimental scaling of Eq.~(\ref{eqn:horacek}), in agreement with our theoretical scaling that depends very weakly on $A$.

As an aside, we note that the theoretical scaling in Eq.~(\ref{eqn:lp_phys}) depends on $q$, $P_\text{SOL}$ and $B_T$ with exponents that are comparable to the ones of the experimental scaling law derived in Ref.~\cite{sieglin2016} from a nonlinear regression performed on $\lambda_q$ measurements of L-mode ASDEX discharges. This nonlinear regression has been carried out by considering the same fitting quantities as the ones considered in the H-mode scaling of Ref.~\cite{eich2011}, providing a link between the L-mode and the H-mode scaling laws. In particular, we note that, combining the dependence on $q$ and $B_T$, the theoretical scaling law of Eq.~(\ref{eqn:lp_phys}) inversely depends on the poloidal magnetic field, a feature shared with the heuristic drift-based H-mode scaling law derived in Ref.~\cite{goldston2011}.

% We start with the scaling based on ASDEX discharges derived in Ref.~\cite{sieglin2016} and reported in Eq.~(\ref{eqn:exp_aug}). The theoretical scaling in Eq.~(\ref{eqn:lp_phys}) shows an almost linear dependence on the safety factor and decreases as $P_\text{SOL}$ and $B_T$ increase, with exponents comparable to the ones in Eq.~(\ref{eqn:exp_aug}). Moreover, it shows a very weak dependence on $A$, again in agreement with the experimental scaling law of Eq.~(\ref{eqn:exp_aug}), which is independent of the ion mass number. Since the experimental scaling law of Eq.~(\ref{eqn:exp_aug}) is derived by using data from a single tokamak, the dependence of $\lambda_q$ on geometrical parameters, such as $R_0$ and $a$, is not included, while the theoretical scaling shows an increase of $L_p$ with both $R_0$ and $a$. On the other hand, the theoretical pressure decay length increases with the edge density, but no density dependence is shown by Eq.~(\ref{eqn:exp_aug}).

The theoretical scaling of Eq.~(\ref{eqn:lp_phys}) with the proportionality constant given by the fitting procedure can be used to predict the SOL width for future tokamaks, such as ITER, COMPASS Upgrade, JT-60SA, and DTT. Considering the baseline scenario just before the L-H transition, one obtains $\lambda_{q,\text{th}} \simeq 3.5$~mm for ITER ($R_0=6.2$~m, $a=2$~m, $q=2$, $P_\text{SOL}=18$~MW, $\kappa = 1.4$, $\bar{n}_e = 4\cdot 10^{19}$~m$^{-3}$, and $B_T=5.3$~T~\cite{artaud2018}), $\lambda_{q,\text{th}} \simeq 1.8$~mm for COMPASS Upgrade ($R_0=0.89$~m, $a=0.27$~m, $q=2.6$, $P_\text{SOL}=3.7$~MW, $\kappa = 1.8$, $\bar{n}_e = 2\cdot 10^{20}$~m$^{-3}$, and $B_T=5.0$~T~\cite{panek2017}), $\lambda_{q,\text{th}} \simeq 7.1$~mm for JT-60SA ($R_0=2.9$~m, $a=1.2$~m, $q=3$, $P_\text{SOL}=10$~MW, $\kappa = 1.9$, $\bar{n}_e = 6.3\cdot 10^{19}$~m$^{-3}$, and $B_T=2.3$~T~\cite{giruzzi2017}), and $\lambda_{q,\text{th}} \simeq 3.0$~mm for DTT ($R_0=2.1$~m, $a=0.6$~m, $q=3$, $P_\text{SOL}=15$~MW, $\kappa = 1.7$, $\bar{n}_e = 1.8\cdot 10^{20}$~m$^{-3}$, and $B_T=6.0$~T~\cite{contessa2019}).
The theoretical value of $\lambda_q$ for ITER L-mode is within the range of values predicted by the experimental scaling laws derived in Ref.~\cite{horacek2020}.

\subsection{Far SOL}

\begin{figure}
    \centering
    \includegraphics[scale=0.85]{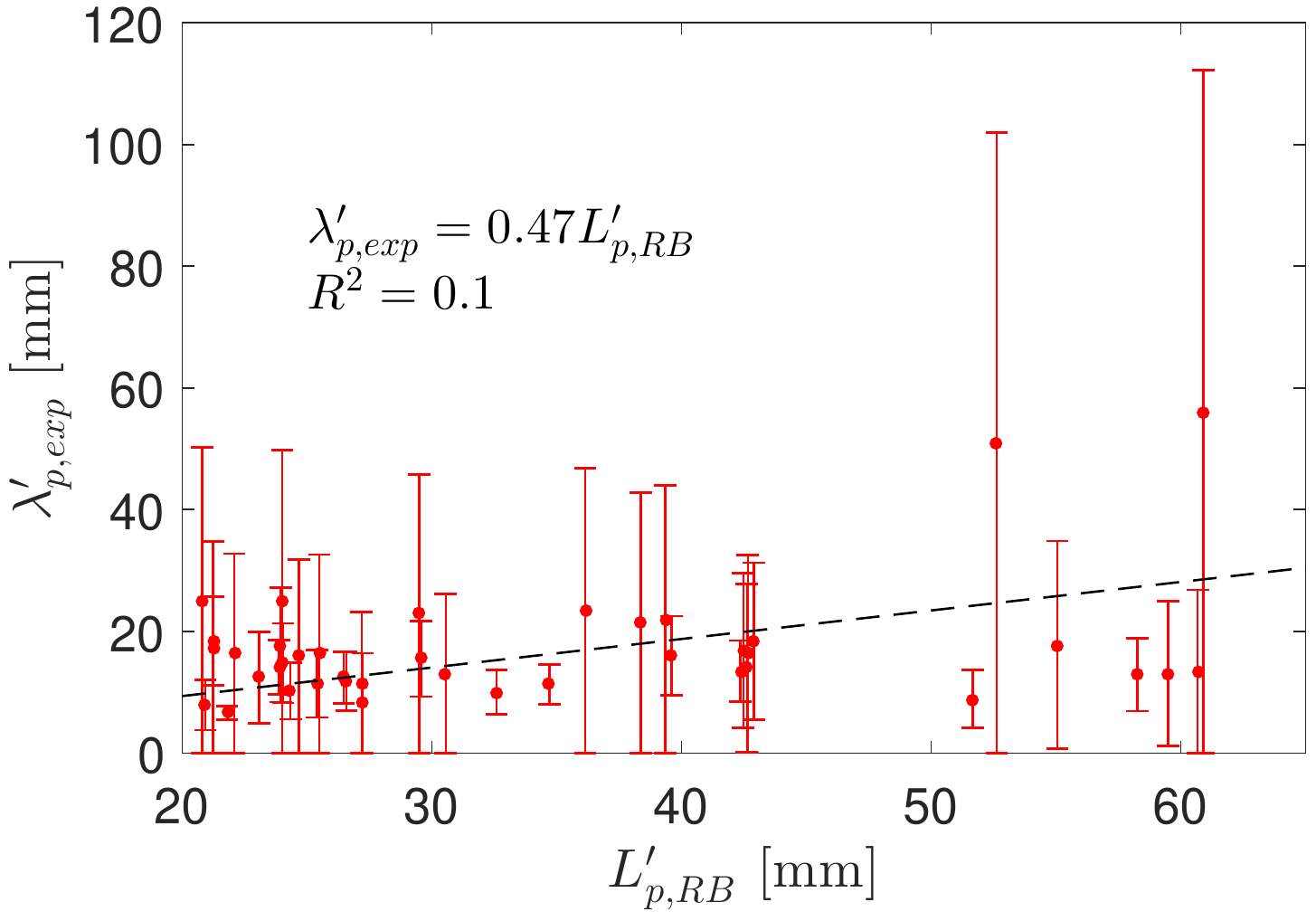}
    \caption{Comparison of the theoretical scaling law of Eq.~(\ref{eqn:far_rb_phys}) to experimental values of pressure decay length measured with a fast reciprocating probe at the outboard midplane of TCV L-mode discharges in conduction regime. Experimental data are taken from Ref.~\cite{tsui2018}. The dashed black line represents the best fit $\lambda_{p,\text{exp}}'= \alpha L_{p,\text{RB}}'$ with $\alpha$ the unique fitting parameter.}
    \label{fig:comp_exp_far}
\end{figure}

The absence of a multi-machine database or experimental scaling laws for the pressure decay length in the far SOL strongly limits the possibility to carry out a complete validation of our theoretical scaling. As a preliminary comparison with experimental data, we consider a set of measurements of the far SOL decay length taken at the outboard midplane of TCV L-mode discharges in lower single-null configuration by using a fast reciprocating probe~\cite{boedo2009fast}.   
Experimental far SOL decay lengths are measured at fixed $B_T=1.4$~T, in both reversed and forward magnetic field direction, at various values of density, plasma current, and connection length. A detailed description of the considered database as well as of the experimental setup are reported in Refs.~\cite{tsui2018,vianello2017}.
The result of the comparison is shown in Fig.~\ref{fig:comp_exp_far}.

There are two main difficulties that affect the fitting of the experimental pressure profiles to derive the far SOL experimental pressure decay length. First, in low-density discharges, the transition between the near and far SOL appears to be very close to the LFS tokamak wall, thus making it difficult to clearly distinguish its position and reducing the numbers of data points available for the fit of the far SOL pressure profile. 
Indeed, as shown in Refs.~\cite{labombard2001,militello2016scrape,kube2018}, the position of the near-far SOL interface depends on the density and moves towards the first wall as the density decreases. 
Therefore, an exponential function with a value of decay length close to the one in the near SOL is able to fit the entire experimental profile and the fit with two exponential functions returns near and far decay lengths that are very similar. 
The discharges that allow a more precise analysis of the far SOL decay length are at high density, in the conduction regime with the presence of significant electron temperature variation along the parallel direction, which questions the applicability of the present model, in particular for the absence of the neutral-plasma interaction processes that might affect the far SOL, as experimentally observed in Refs.~\cite{wynn2018,vianello2019}.
For the comparison of the far SOL decay length presented in Fig.~\ref{fig:comp_exp_far}, we choose to exclude the discharges that do not allow a clear identification of the near and far SOL and consider high-density discharges, despite the questions on the applicability of this model.

The second difficulty emerges when fitting experimental data at high value of $\lambda_q'$. Indeed, a small variation of the fitting range produces a large variation of $\lambda_q'$. This is reflected on large experimental uncertainties that prevent us from an accurate comparison with the theoretical prediction and potentially hide some dependencies. 
In fact, as shown by the error propagation, the relative uncertainties of $\lambda_{p,\text{exp}}'$ inversely depend on the radial derivative of the pressure profile, meaning that a particularly flat radial pressure profile leads to large uncertainties of $\lambda_{p,\text{exp}}'$. 

The subset of the database considered for this comparison includes discharges that are mainly in the RB regime~\cite{tsui2018} and hence we fit experimental data by using the theoretical RB scaling law in Eq.~(\ref{eqn:far_rb_phys}) with the unique fitting parameter being the proportionality constant between experimental measurements and $L_{p,\text{RB}}$. The quality of the fit is then expressed through the $R^2$ parameter. 
As shown in Fig.~\ref{fig:comp_exp_far}, there is a very weak correlation between theoretical predictions and experimental data, being $R^2$ only slightly positive.

\section{Conclusions}\label{sec:conclusions}

A theoretical scaling of the pressure and density decay lengths in the near SOL of L-mode diverted plasma discharges valid in low-recycling conditions is analytically derived from an electrostatic two-fluid model by balancing the heat source in the core region, the perpendicular heat flux crossing the separatrix, and the parallel losses at the vessel walls. 
Similarly, by balancing the perpendicular turbulent transport due to plasma filament motion and the parallel flow, the far SOL pressure and density exponential decay lengths are analytically derived in the RB and RX filament regimes.

The theoretical scaling laws for pressure and density decay lengths in the near and far SOL are then compared to the results of flux-driven, global, two-fluid turbulent simulations in a lower single-null geometry, carried out by using the GBS code. 
In the near SOL, there is a very good agreement between theoretical and numerical results. Indeed, across the entire set of simulations considered in this work, the difference between simulation results and theoretical predictions is below 20\%. 
In the far SOL, a pattern-recognition algorithm for filament detection/tracking is applied to the simulation results to determine filament size and velocity, and to identify the filament motion regime. Detected filaments in our simulations mainly belong to the RB and RX regimes. 
The theoretical estimates of the far SOL pressure and density decay lengths in RB and RX regimes agree with simulation results within an error of 20~\% for the pressure and 40~\% for the density, pointing out that the model considered here contains the main physics, although the dispersion of simulation results around the analytical prediction suggests the need of future investigations  with a more accurate model for the filament velocity, which accounts for the filament-filament interaction and filament rotation.

A comparison between the theoretical scaling of the pressure decay length in the near SOL and experimental measurements of the power fall-off length, taken from the multi-machine database presented in Ref.~\cite{horacek2020} and extended by adding TCV data from Ref.~\cite{maurizio2017}, is carried out with the only fitting parameter being the proportionality constant between the power fall-off length and the near SOL pressure decay length. Our model reproduces experimental data with a very high value of the quality parameter, $R^2\simeq 0.85$. 
The theoretical scaling with the proportionality constant from the fit predicts a near SOL width for ITER L-mode plasma of 3.5~mm, a value close to the one predicted by the experimental scaling laws derived in Ref.~\cite{horacek2020}. 

Analogously, the theoretical scaling law of the far SOL pressure decay length in the RB regime is compared to experimental measurements obtained from a fast reciprocating probe located at the outboard midplane in TCV L-mode lower single-null discharges~\cite{tsui2018}. 
This preliminary comparison shows a very weak correlation between theoretical prediction and experimental results ($R^2\simeq 0.1$).
However, the presence of significant temperature variation along the magnetic field lines in the far SOL observed experimentally for the considered discharges and not included in the present work, the experimental difficulty in identifying at low density the interface between near and far SOL, and the large experimental uncertainties affecting measurements of long pressure decay lengths make the present comparison not conclusive and prompt for the need of further investigations in high-recycling conditions and possibly including data from multiple tokamaks and diagnostics.
In addition, the model considered here should be generalised to include the effect of neutrals and impurities, that we expect to have an important role in the formation of the far SOL profile.

\section*{Acknowledgments}
The authors thank C. F. Beadle and C. Theiler for useful discussions.
The simulations presented herein were carried out in part at the Swiss National Supercomputing Center (CSCS) under the project ID s882 and in part on the CINECA Marconi supercomputer under the GBSedge project. 
This work, supported in part by the Swiss National Science Foundation, was carried out within the framework of the EUROfusion Consortium and has received funding from the Euratom research and training programme 2014 - 2018 and 2019 - 2020 under grant agreement No 633053. The views and opinions expressed herein do not necessarily reflect those of the European Commission. This work was supported in part by the US Department of Energy under Award Number DE-SC0010529 and by MEY S projects \#LM2018117 and CZ.02.1.01/0.0/0.0/16\_013/0001551.

\section*{References}
\bibliographystyle{unsrt}
\bibliography{library}

\end{document}